\documentclass[fleqn,usenatbib]{mnras}
\usepackage{float}
\usepackage{newtxtext,newtxmath}

\usepackage[T1]{fontenc}

\DeclareRobustCommand{\VAN}[3]{#2}
\let\VANthebibliography\thebibliography
\def\thebibliography{\DeclareRobustCommand{\VAN}[3]{##3}\VANthebibliography}

\usepackage{graphicx}	
\usepackage{amsmath}	

\title[WHIM Absorption with CAMELS]{X-ray Absorption Lines in the Warm-Hot Intergalactic Medium: Probing {\em Chandra} observations with the CAMEL simulations}

\author[Butler Contreras et al.]{Amanda Butler Contreras$^{1,2}$\thanks{E-mail: amanda.butler@aya.yale.edu},
Erwin T. Lau$^{3,4}$\thanks{E-mail: erwin.lau@cfa.harvard.edu},
Benjamin D. Oppenheimer$^{5}$,
\'Akos Bogd\'an$^{3}$, 
Megan Tillman$^{6}$,
\newauthor
Daisuke Nagai$^{1,2}$,
Orsolya E. Kov\'acs$^{7}$,
Blakesley Burkhart$^{6,8}$
\\
\\
$^{1}$Department of Physics, Yale University, New Haven, CT 06520, USA \\
$^{2}$Department of Astronomy, Yale University, New Haven, CT 06520, USA \\
$^{3}$Center for Astrophysics | Harvard \& Smithsonian, 60 Garden Street, Cambridge, MA, 02138, USA \\
$^{4}$Department of Physics, University of Miami, Coral Gables, FL 33124, USA\\
$^{5}$CASA, Department of Astrophysical and Planetary Sciences, University of Colorado, 389 UCB, Boulder, CO 80309, USA\\
$^{6}$ Department of Physics and Astronomy,
Rutgers University, Piscataway, NJ 08854, USA\\
$^{7}$Department of Theoretical Physics and Astrophysics, Faculty of Science, Masaryk University, Kotl\'a\v{r}sk\'a 2, Brno, 611 37, Czech Republic\\
$^{8}$ Center for Computational Astrophysics, Flatiron Institute, 162 Fifth Avenue, New York, NY 10010, USA \\
}

\pubyear{2023}

\begin{document}
\label{firstpage}
\pagerange{\pageref{firstpage}--\pageref{lastpage}}
\maketitle

\begin{abstract}
Known as the ``Missing Baryon Problem'', about one-third of baryons in the local universe remain unaccounted for. The missing baryons are thought to reside in the warm-hot intergalactic medium (WHIM) of the cosmic web filaments, which are challenging to detect.
Recent \textit{Chandra} X-ray observations  used a novel stacking analysis and detected an OVII absorption line toward the sightline of a luminous quasar, hinting that the missing baryons may reside in the WHIM. 
To explore how the properties of the OVII absorption line depend on feedback physics, we compare the observational results with predictions obtained from the Cosmology and Astrophysics with MachinE Learning (CAMEL) Simulation suite. CAMELS consists of cosmological simulations with state-of-the-art supernova (SN) and active galactic nuclei (AGN) feedback models from the IllustrisTNG and SIMBA simulations, with varying strengths. We find that the simulated OVII column densities are higher in the outskirts of galaxies than in the large-scale WHIM, but they are consistently lower than those obtained in the \textit{Chandra} observations, for all feedback runs. We establish that the OVII distribution is primarily sensitive to changes in the SN feedback prescription, whereas changes in the AGN feedback prescription have minimal impact. We also find significant differences in the OVII column densities between the IllustrisTNG and SIMBA runs. 
We conclude that the tension between the observed and simulated OVII column densities cannot be explained by the wide range of feedback models implemented in CAMELS. 
\end{abstract}




\section{Introduction}
In the current standard $\Lambda$CDM cosmological model, baryons only make up a small part of the total energy and matter content of the Universe. Big Bang Nucleosynthesis \citep[BBN,][]{cooke2014} and studies of the Cosmic Microwave Background (CMB) have established the cosmic baryon mass density to be $\Omega_b = 0.0449$ \citep{plancketal}. Observations of distributions of galaxies in the local Universe also revealed a cosmic web pattern of the Universe's large-scale structure \citep[e.g.][]{geller_huchra89}, consistent with our current understanding of structure formation \citep[e.g.][]{millenium}. 

However, measurements of the baryon content of the local low-redshift $z<2$ Universe do not corroborate with the BBN and CMB studies of the baryon budget. Only $60\%$ to $70\%$ of expected baryons have been uncovered \citep{shull12}. This is widely known as the ``Missing Baryon Problem'' \citep{copi95, fukugita_etal98, burles01,bregman07}.

Despite the challenges, there have been several attempts at detecting WHIM. Light from background quasars can be absorbed by the WHIM, producing absorption line features. Absorption lines of Ly$\alpha$ and OVI in the ultraviolet (UV) range have been detected around several quasars, providing evidence of the existence of the warm phases of the WHIM \citep[e.g.,][]{bahcall93,jannuzi98,tripp98,tripp00,tripp08,lehner07,stocke14,danforth16}. Because UV observations are only sensitive to the lower temperature WHIM, a substantial fraction of the hot ($T>10^5$~K) phases of the WHIM remain undetected \citep{smith11,oppenheimer12,rahmati16}. 

The bulk of the WHIM at higher temperatures is expected to emit weakly in the X-ray wavelengths. While X-ray emission has been detected from the hottest and highest density WHIM in the outskirts of galaxy clusters \citep{werner08,eckert15,alvarez18,tanimura20}, the density and gas temperature  of the dominant fraction of the WHIM is too low to be detected in emission or individual absorption systems. Previous studies were highly controversial \citep{nicastro05, kaastra06} and/or suffered from low detection significance \citep[e.g.,][]{mathur03}, demonstrating the challenges of detecting WHIM. Some of the absorption signal may also come from within halos of galaxies or galaxy groups instead of the large-scale WHIM \citep[e.g.,][]{nicastro18,johnson19,dorigo_jones_etal22}. Additionally, these detections were obtained without {\em a priori} information of the redshifts of the detection, making it even more difficult to ascertain the origins of the detections as WHIM in the local Universe.  

Recently, \citet{kovacs19} (hereafter K19) used 17 HI Lyman-$\alpha$ absorbers with known redshifts that are associated with galaxies \citep{tripp98} to enhance the signal-to-noise ratio of OVII absorption lines in the X-ray spectrum of the background quasar, H1821+643. This method used the redshifts of the HI absorbers to shift the OVII lines measured with \textit{Chandra}/LETG to the rest frame. With the spectra shifted into the rest frame for each line, they were co-added to enhance the signal-to-noise. This process revealed an OVII absorption line feature at a rest frame wavelength of 21.6~\AA, with a statistical significance of $3.3\sigma$. A column density of $N_\mathrm{OVII} = (1.4 \pm 0.4) \times 10^{15} \ \mathrm{cm^{-2}}$ was computed, which is translated to an overall WHIM contribution of ($37.5 \pm 10.5$)\% to the total cosmic baryon mass density. 
To understand whether the OVII absorption line detected in K19 originates from the outskirts of galactic haloes or from the cosmic web filament, and how feedback from galaxies affect the WHIM signals, we compare the observational results with those predicted by cosmological simulations. To this end, we use the novel Cosmology and Astrophysics with MachinE Learning (CAMEL) simulation suite \citep{camelsrelease}. The CAMEL simulation suite is a unique set of cosmological hydrodynamical simulations with varying supernovae (SN) and active galactic nuclei (AGN) feedback.
The CAMEL simulations use two galaxy formation hydrodynamic simulation codes as their fiducial models: IllustrisTNG \citep{pillepich18, nelson18} and SIMBA \citep{dave19}, and then vary SN and AGN feedback prescriptions in a series of smaller volume simulations that allow one to examine how the parameter dependences of feedback physics impact the WHIM properties.  We therefore aim to determine if a particular set of feedback parameters can recreate the purported K19 detection.  

We organize the paper as follows.  In Section~\ref{methods} we give an overview of the CAMEL simulations, and also describe the methods for generating column densities, mocking ``stacked'' column densities for comparison with K19, and our data analysis.  We present our results in Section~\ref{results}, with particular attention to splitting our sample based on distance to the nearest galaxy, as well as determining how the OVII column density strength relates to SN and AGN feedback.  Finally, in Section~\ref{summary} we give a summary and discussion of our results, as well as outline the next steps in the project.

\section{Methods}\label{methods}

\subsection{CAMEL Simulations} \label{2.1}
The Cosmology and Astrophysics with MachinE Learning Simulations (CAMELS) dataset \citep{camelsrelease} is a robust set of 4,233 cosmological simulations, 2,049 N-body, and 2,184 hydrodynamical simulations. The simulations are based on a $\Lambda$CDM cosmological model and span a wide range of data objects ranging from galaxy halos to spectra and radial profiles. All simulations consist of a $(25 h^{-1}\mathrm{\,comoving\,Mpc})^3$ volume with base cosmological parameters $\Omega_b = 0.049$, $h = 0.6711$, and $n_s = 0.9624$. The simulations are additionally based on the Illustris TNG and SIMBA simulation models, run using the AREPO and GIZMO codes, respectively; each of these models can be explored separately, as we do in this project. Additionally available in CAMELS are snapshots at redshifts ranging from $z = 6$ to $z = 0$, and we explore simulations at $z = 0.154$, to match the median redshift of the HI absorbers in K19. We use the 1P subset of CAMELS, a subset of 61 simulations that vary one cosmological or astrophysical parameter at a time across a range of 11 values (including the fiducial, or base case value) for each. These parameters include the cosmological parameters $\Omega_m$ and $\sigma_8$, and four astrophysical parameters, two corresponding to supernovae feedback and the other two to AGN feedback. We do not explore the effects of cosmology in this project, thus using a reduced 1P set. It is important to note that the SIMBA and IllustrisTNG simulations implement the astrophysical feedback differently. In addition, SIMBA tracks dust grains, and IllustrisTNG includes magnetohydrodynamics. The gravity and hydrodynamics solvers are different between the IllustrisTNG \citep{pillepich18, nelson18} and SIMBA suites \citep{dave19}. 

The four modes of SN and AGN feedback are parameterised as $A_{\mathrm{SN1}}$, $A_{\mathrm{SN2}}$,  $A_{\mathrm{AGN1}}$, $A_{\mathrm{AGN2}}$.
In particular, the parameter $A_{\mathrm{SN1}}$ represents a normalisation factor of the galactic wind feedback flux. This is implemented as either a pre-factor for the overall energy output per unit star formation (IllustrisTNG) or as a pre-factor for the wind mass outflow rate per unit SFR (SIMBA). As $A_{\mathrm{SN2}}$ represents the normalisation factor for the galactic wind speed, varying $A_{\mathrm{SN2}}$ in IllustrisTNG maintains fixed energy output through adjustment of both wind speed and mass loading factor. In SIMBA, the mass loading factor stays fixed, and $A_{\mathrm{SN2}}$ varies the wind speed jointly with wind energy flux. As for the AGN parameters, $A_{\mathrm{AGN1}}$ is a normalisation factor for the energy output of AGN feedback. In IllustrisTNG, this is implemented as the pre-factor power in kinetic feedback, while in SIMBA, $A_{\mathrm{AGN1}}$ is the pre-factor for the momentum flux of mechanical outflows in quasar and jet feedback. $A_{\mathrm{AGN2}}$ does not have a common definition across the two simulations but is defined as the parameterisation of heated gas temperature and ``burstiness'' in AGN bursts (IllustrisTNG) and an adjustment to the speed of continuously-driven AGN jets (SIMBA). The ranges of variation for $A_{\mathrm{SN1}}$ and $A_{\mathrm{AGN1}}$ is $[0.25, 4.0]$, while for $A_{\mathrm{SN2}}$ and $A_{\mathrm{AGN2}}$ it is $[0.5, 2.0]$. 
It is important to note that in contrast to the aforementioned astrophysical parameters, the cosmology parameter effects are designed to be the same across simulations. 

\subsection{2D CAMELS Column Density Maps} \label{2.2}
\begin{figure*}
    \includegraphics[width=0.9\textwidth]{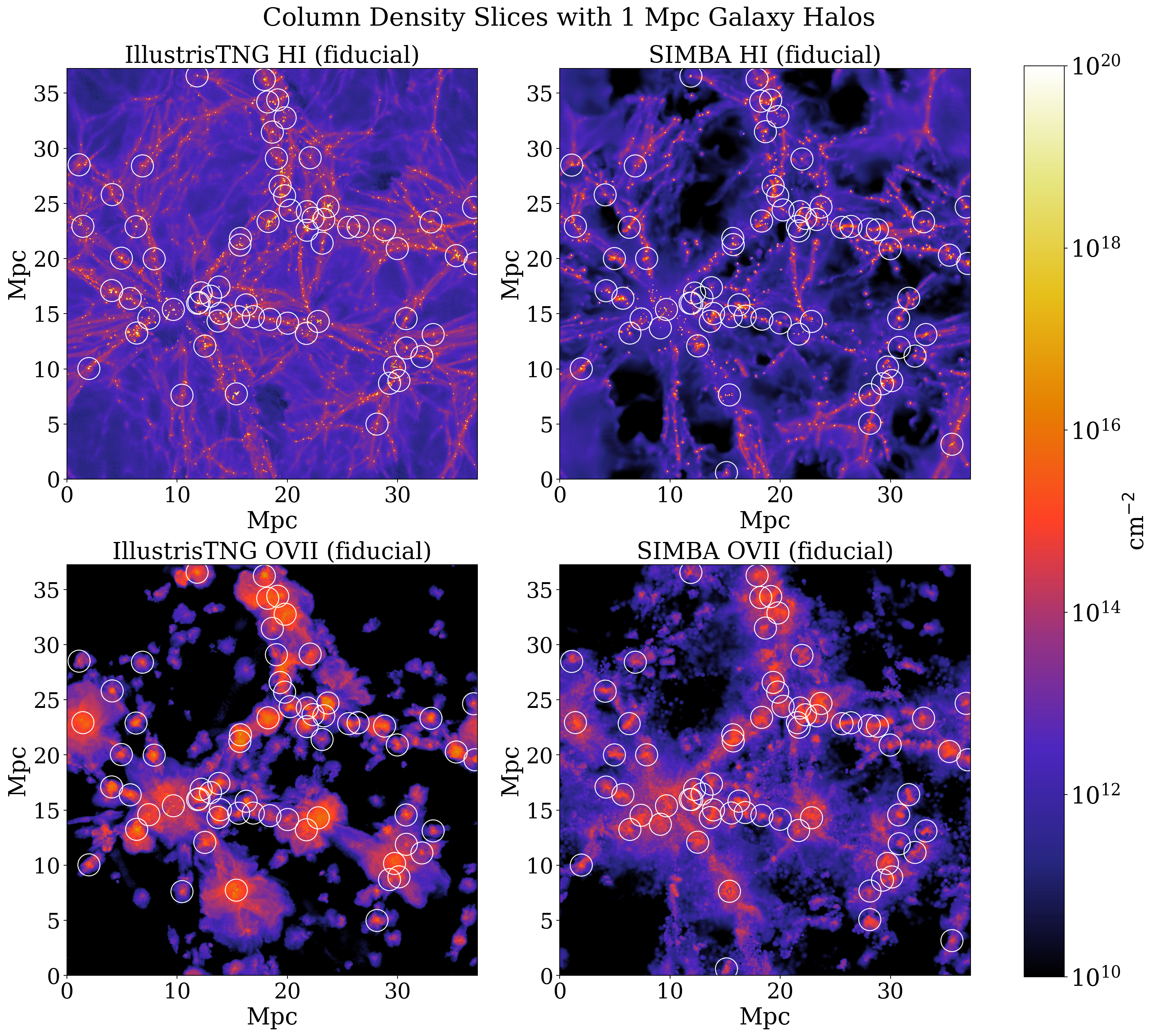}
    \caption{HI (top panels) and OVII (bottom panels) column density maps from the fiducial runs of the IllustrisTNG (left) and SIMBA (right) 1P simulation at $z=0.154$ from a slice with width $4.167\,h^{-1}$ Mpc. The colour indicates the value of the column density in units of cm$^{-2}$. Overplotted are white 1~Mpc halo radii around galaxies in the simulation, which we define to be the halo outskirts of the galaxies. 
    The halos are selected to have virial masses $M_{\rm vir} \geq 3\times 10^{11} \ {\rm M}_{\odot}$, and with their centres lying within the slice. 
    Note that this visualisation does not represent the methods of selecting regions for masked pixel maps described in Section~\ref{2.5}.}
    \label{fig1}
\end{figure*}

For this project, we use CAMELS to generate 2D projected numerical column density maps for the OVII and HI absorbers. We use the Trident code \citep{hummels2017} to generate pixel column density maps in HI and OVII. To calculate the column densities, Trident interpolates fluid elements onto 2D grids of a slice using a line-of-sight (LOS) integral for an ion density field $f(x)$ along LOS $\hat{n}$:
\begin{equation}
    g(X) = \int f(x)\;\hat{n}\cdot dx. \label{eq:unweighted}
\end{equation}
Figure~\ref{fig1} presents an example of the 2D column density maps for OVII and HI for the fiducial simulations. We create $4000\times 4000$ pixel grids, which translates to a pixel size of $9.3$~kpc at $z=0$. We additionally use six slices, each of one-sixth the depth of the box ($4.167\,h^{-1} \mathrm{comoving\,Mpc}$ per slice) to create our maps, which are deemed appropriate for the depth of Lyman-$\alpha$ absorber objects at approximately 400~km~s$^{-1}$ \citep{Wijers}. Each snapshot thus returns a total of $9.6\times 10^7$ pixel column densities for both HI and OVII.  

When calculating the simulated column densities, we account for the main ionising processes that determine the ionisation balance for these two ions. Photoionisation by the extragalactic ultraviolet background (UVB) radiation dominates for HI, while collisional ionisation is dominant for OVII. We recalibrate the strength of the UVB to reproduce the observed statistics of the Lyman-$\alpha$ Forest, which requires a UVB photoionisation correction for each simulation run as described in Appendix~\ref{sec:uv_correction}. 

\subsection{HI Column Densities from Observation} \label{2.3}

To match the simulated HI column densities with those used in K19, we convert the equivalent widths (EW) of their Lyman-$\alpha$ lines to HI column densities ($\mathrm{N_{HI}}$). In the optically-thin regime, we can use the following conversion: 
\begin{equation}
    N_\mathrm{HI} = 1.13\times 10^{17}\mathrm{cm^{-2}} \frac{\mathrm{EW}}{\lambda^2f_\mathrm{osc}},
\end{equation}
where $f_\mathrm{osc} = 0.4164$ is the HI absorption oscillator strength, and $\lambda = 1215.67$~\AA~is the restframe wavelength of the Lyman-$\alpha$ line ~\citep{verner}. Table~\ref{tab:tab2} lists the redshifts, equivalent widths, and calculated column densities of the Lyman-$\alpha$ absorption lines in K19. 

\begin{table}
	\centering
	\caption{Summary of redshifts, equivalent widths in m\AA, and column densities $N_\mathrm{{HI}}$ for each of the 17 Lyman-$\alpha$ absorption lines of the H1821+643 quasar spectra in K19 and \citep{tripp98}.}
	\label{tab:tab2}
	\begin{tabular}{ccc} 
		\hline
		 $z$ & equivalent width (mÅ) & 0.1 dex bin [$\mathrm{log_{10}}$($\mathrm{N_{HI}/\mathrm{cm^{-2}}})$]\\
		\hline
    
		0.05704 & 87  & 13.2 \\
		0.06432 & 62  & 13.0\\
		0.08910 & 47 & 12.9\\
		0.11152 & 66 & 13.0\\
		0.11974 & 102 & 13.2\\
		0.12157 & 353 & 13.8 \\
		0.12385 & 35 & 12.8\\
		0.14760 & 229 & 13.6\\
        0.16990 & 523 & 13.9\\
        0.18049 & 75 &  13.1\\
        0.19905 & 29 & 12.7\\
        0.22489 & 739 & 14.1\\
        0.24132 & 79 & 13.1\\
        0.24514 & 79 & 13.1\\
        0.25814 & 134 & 13.3\\
        0.26156 & 163 & 13.4\\
        0.26660 & 163 & 13.4\\
		\hline
    \end{tabular}
\end{table}

\subsection{CDDF and Stacking Analysis} \label{2.4}

We calculate the OVII and HI Column Density Distribution Function (CDDF), defined as ${d^2n}/({d\log_{10} N dz})$ (where $n$ is the number of absorbers and $N$ is the column density), for the range of column density values $\log_{10}(N_\mathrm{OVII} /\mathrm{cm^{-2}}) \in [5,20]$ and $\log_{10}(N_\mathrm{HI} /\mathrm{cm^{-2}}) \in [10,25]$ with bin width of 0.1 dex, or 150 bins, which matches the precision of the observation in K19. We calculate $n$ as the number of pixel counts in a $0.1$ dex bin, $d\log_{10} N = 0.1$, and with redshift pathlength $dz = 0.0015$ (for a slice of full volume with $dz = 0.0090$), which we scale by the total number of pixels $N_{\rm pixel} = 9.6 \times 10^7$.

We then examine the CDDF of OVII at specific values of $N_\mathrm{{HI}}$ corresponding to those of the 17 systems in the observation data in K19. 
Our $0.1$~dex binning leads to duplicates in the HI column density values. For the 17 systems, there are 12 distinct values of $N_\mathrm{{HI}}$. 

We also split the sightlines into two groups based on whether they were classified as corresponding to galaxy halo outskirts or cosmic web filament Lyman-$\alpha$ absorbers in K19 (see Section \ref{2.5} for more details on this separation criteria). 
This results in the total number of 15 distinct values of $N_\mathrm{{HI}}$, with 8 corresponding to the filamentary gas, and 7 for the extended halo regions. While there is overlap in the ranges of $N_\mathrm{{HI}}$ values between these groups, we note that the highest values are in halo regions and the lowest are in filaments. This phenomenon is visible in Figure~\ref{fig2} and Figure~\ref{fig4}.

We then isolate $N_\mathrm{OVII}$ distributions for each Lyman-$\alpha$ absorber sightline at its corresponding $N_\mathrm{OVII}$ bin and calculate the mean of each distribution as a proxy for the observed $N_\mathrm{OVII}$ at each sightline. 
By averaging these values for a single simulation and feedback parameter, we can estimate the ``stacked'' value of $N_\mathrm{OVII}$ for that cosmology configuration. 

\subsection{Gas in Outskirts of Galactic Haloes and in Cosmic Web Filaments}
\label{2.5}

The OVII column density ($N_\mathrm{OVII}$) can originate from WHIM in the cosmic web filament or gas in the outskirts of galaxy haloes. Physically, we expect the gas nearer to the haloes to be denser and more easily affected by the feedback process in the halo. 
Following K19, we split the absorbers into two samples based on their impact parameter from their nearest galaxies with $ M_{\rm vir} \geq 3\times 10^{11} \ {\rm M}_{\odot}$. Those with impact parameter less than $b < 1$~Mpc are referred to as `halo outskirts', and those with $b > 1$~Mpc are referred to as `cosmic web filaments'.
Note that we refrain from referring to the halo outskirt gas as the circumgalactic medium (CGM), as the CGM commonly refers to the gas within the virial radii of the galactic haloes, which in our case are smaller than 1~Mpc. Also note the parameter $b = 1$~Mpc criteria was chosen in K19 simply because it resulted in an approximately equal number of sightlines in both bins. To approximate the K19 cut, we use the coordinates for the galaxies in the fiducial simulations of SIMBA and IllustrisTNG and consider any projected absorber in the depth of the simulation volume within 1 Mpc as halo outskirt gas. K19 allowed velocity differences between the absorber and galaxy to exceed 1000~km~s$^{-1}$. Thus, we try to mimic this separation by allowing galaxies in all six of our slices to be associated with any absorber within 1~Mpc. The white circles in Figure~\ref{fig1} indicate the 1~Mpc circular regions surrounding the galactic halos for both IllustrisTNG and SIMBA fiducial runs. In most regions, the OVII absorbers with column densities $\log_{10} (N_\mathrm{OVII}/\mathrm{cm^{-2}}) > 14$ fall within 1~Mpc around galactic halos. 
We apply these galaxy positions to other simulations with varied feedback parameters assuming that galaxies remain in roughly equivalent positions, even though their stellar masses change.  

To more accurately explore true CGM regions, we modify the procedure above to select $b < 300$~kpc circular regions around galaxies with $ M_{\rm vir} \geq 3\times10^{11} \ {\rm M}_{\odot}$ as `CGM'. For this selection, we also limit our galaxies by whether their central coordinates are located within a particular slice of the box, rather than looking at all galaxies within all six slices of the full box volume. We look at the full volume (six slices) to compile our statistics. 
\section{Results}\label{results}

\begin{figure}
    \centering
    \includegraphics[width=0.48\textwidth]{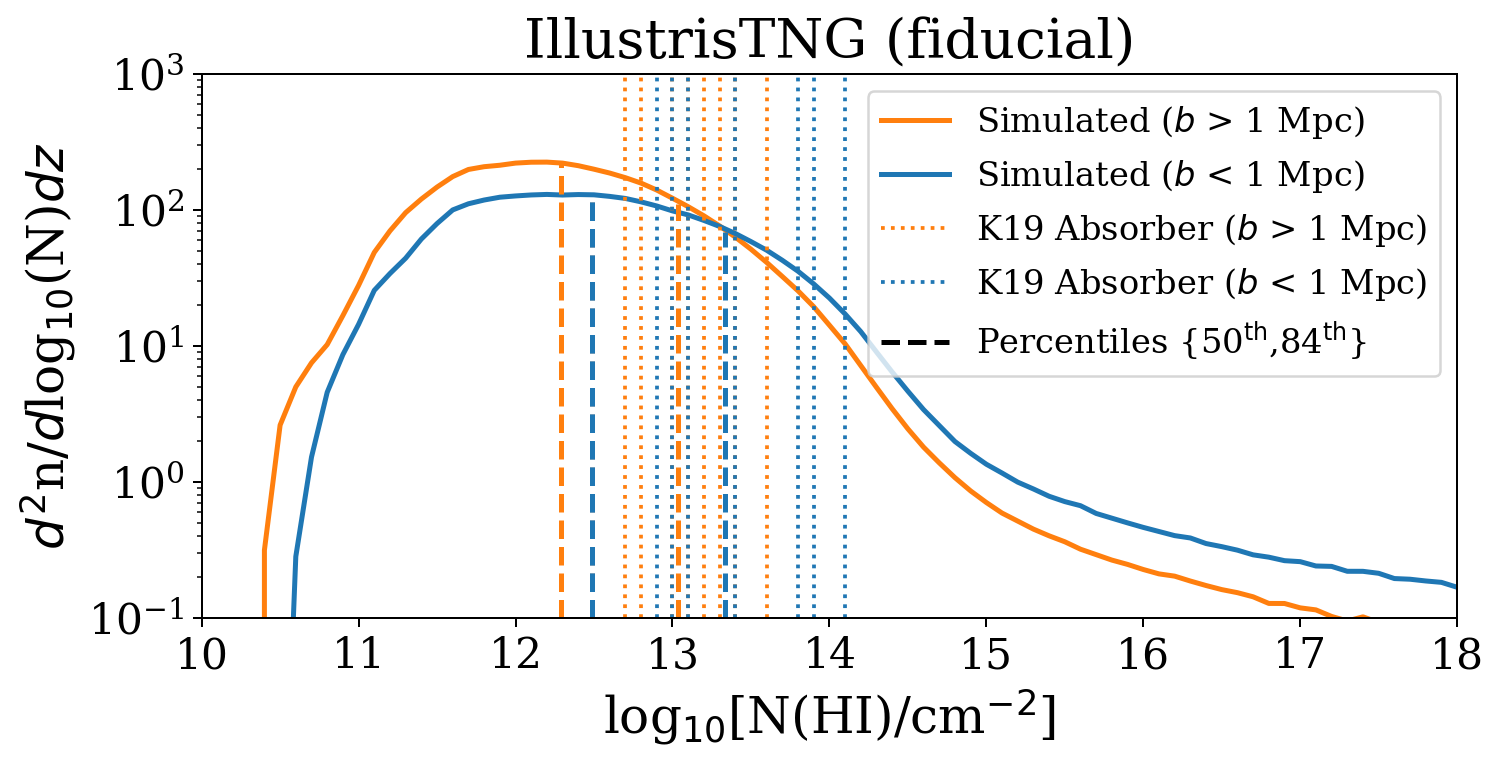}
    \includegraphics[width=0.48\textwidth]{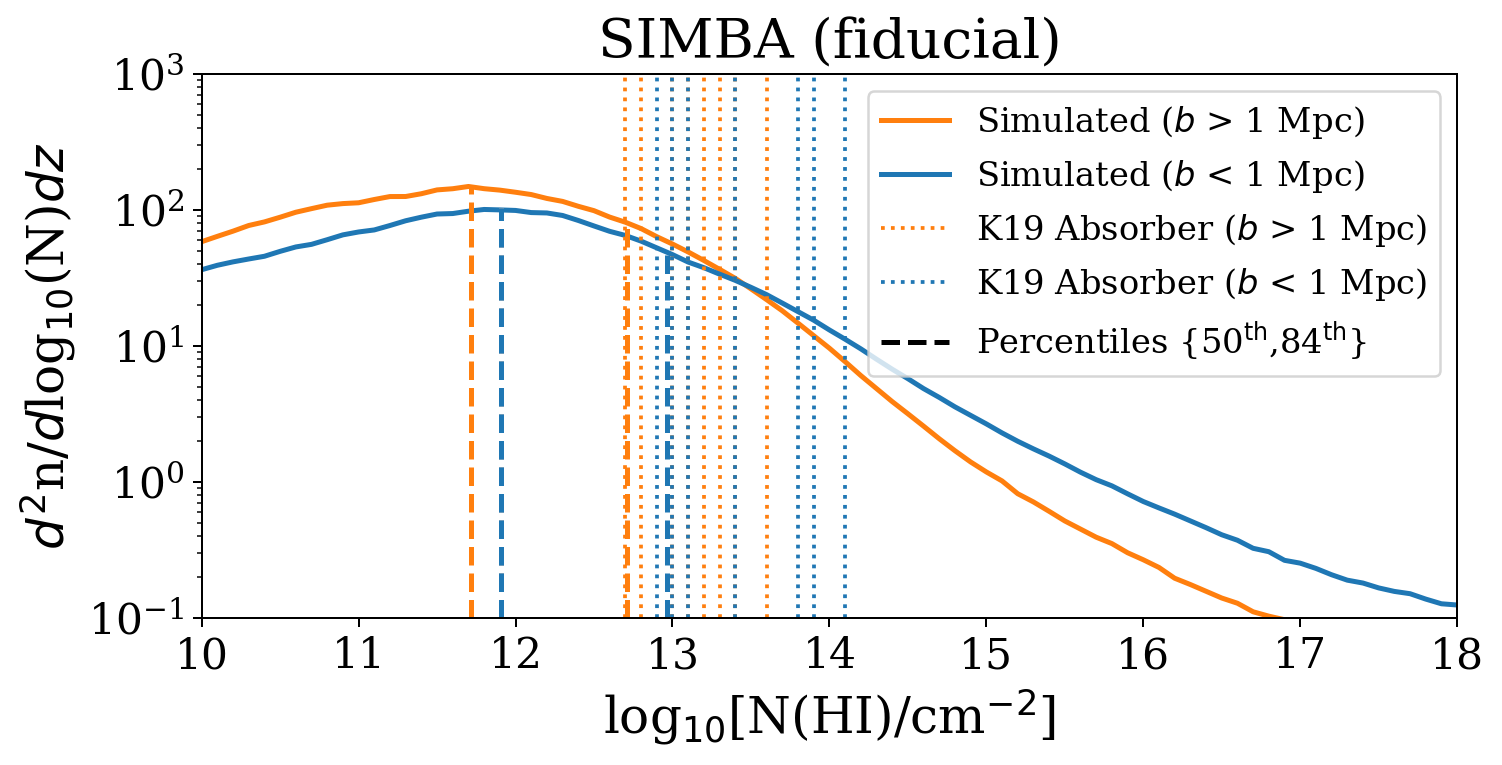}
    \caption{Column Density Distribution Functions for the binned $N_\mathrm{HI}$ data in IllustrisTNG (top) and SIMBA 
    (bottom) separated by the filament or halo outskirt pixel masking, based on the impact parameter $b = 1$~Mpc from galaxy halos. The 50th (median) and 84th percentiles of these distributions are overplotted in dashed lines. From these results, we observe a slight shift in the medians, where the outskirt-only distributions have higher median $N_\mathrm{HI}$ values. The distributions indicating the halo outskirts ($b < 1$~Mpc) are shifted towards higher $N_\mathrm{HI}$ values than the filament ($b > 1$~Mpc) distributions, with smaller peaks in lower $N_\mathrm{HI}$ regimes and a greater abundance in higher $N_\mathrm{HI}$ values. These findings agree with our expectation that halo outskirt regions will contain more concentrated $N_\mathrm{HI}$ than WHIM filaments. Additionally, we overplot the lines corresponding to HI absorbers in K19 (dotted), indicated by whether they are filament (orange) or halo outskirt (blue). The shift from the filament to halo outskirt here reflects the shift in the median. 
    The K19 $N_\mathrm{HI}$ values are larger than the median of the IllustrisTNG distribution, and larger than the 84th percentile of the SIMBA distribution. }
    \label{fig2}
\end{figure}

\begin{figure}
\centering
\includegraphics[width=0.48\textwidth]{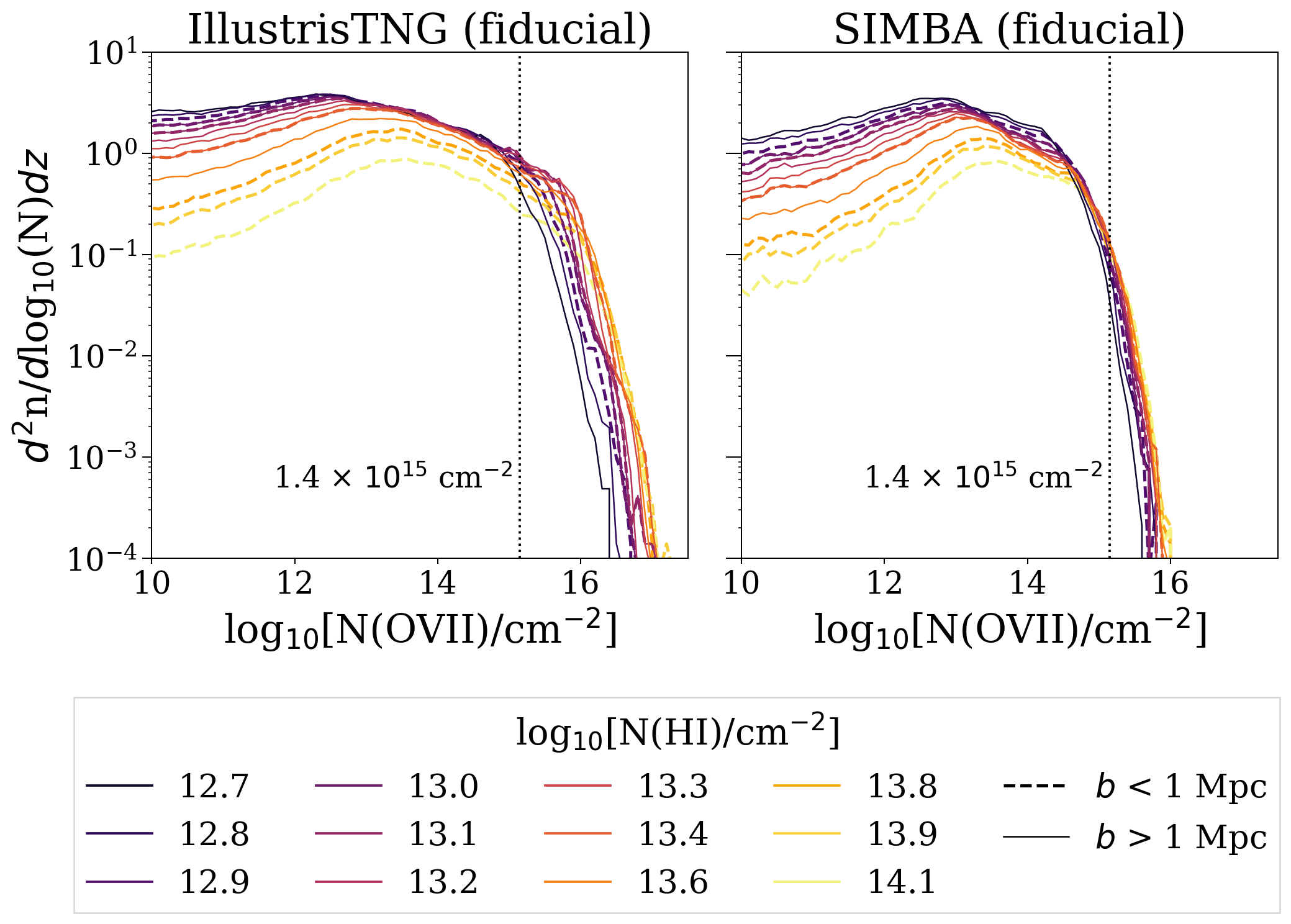}
\caption{Distributions of $\log_{10} (N_\mathrm{OVII}/\mathrm{cm^{-2}})$ at each HI column density corresponding to the 8 filament HI absorbers and 7 halo outskirt HI absorbers from K19 for IllustrisTNG (left) and SIMBA (right) fiducial runs. The colour of the distribution line indicates the value of the $\log_{10} (N_\mathrm{HI}/\mathrm{cm^{-2}})$ sightline the OVII corresponds to, while the dashed lines indicate the K19 halo outskirt absorbers and solid lines indicate filament absorbers. Similarly, overplotted is the result from K19 (vertical, black dotted line). The higher $N_\mathrm{HI}$ values correspond to lower counts of $N_\mathrm{OVII}$ values, up until $\log_{10}(N_\mathrm{OVII}/\mathrm{cm^{-2}}) \approx 15$, where the opposite becomes true. The distribution shapes between SIMBA and IllustrisTNG vary from one another, especially once this aforementioned threshold is crossed, with more high OVII values ($\log_{10}(N_\mathrm{OVII}/\mathrm{cm^{-2}}) \ge 16$) being accounted for in IllustrisTNG than SIMBA.}
\label{fig3}
\end{figure}

\subsection{HI Column Density Distributions}\label{3.1}

In Figure~\ref{fig2}, we show the 
Column Density Distribution Functions for the binned $N_\mathrm{HI}$ data in SIMBA and IllustrisTNG for both filaments and halo outskirts. In units of $\log_{10}( N_\mathrm{HI}/\mathrm{cm^{-2}})$, the median HI CDDF of the filament component in SIMBA is $11.72$, and $11.90$ for the halo outskirt component; for the IllustrisTNG, the median HI CDDF for filament is $12.29$, and for halo outskirt it is $12.48$. This illustrates that the halo outskirts have higher median $N_\mathrm{HI}$ than the filaments. This is in agreement with our expectation and the results from K19 that the denser halo outskirt regions contain more $N_\mathrm{HI}$ absorbers than the filaments. It also shows that the IllustrisTNG predicts higher median $N_\mathrm{HI}$ than SIMBA. 

We also plot the HI column density values from the sightlines in K19 as vertical lines in Figure~\ref{fig2}. To have an appropriate comparison with K19, the sightlines in the simulations are selected using the same criteria in K19 (see Section~\ref{2.5}). The K19 HI column densities are overall higher than the medians of the filament and halo outskirt distributions in the IllustrisTNG simulation and higher than the 84th percentile (roughly corresponding to $1\sigma$ above the median) in the SIMBA simulation. This suggests that the K19 sightlines are probing the denser parts of the filament and halo outskirts. 

It may be unexpected that Figures~\ref{fig1} and \ref{fig2} show such large differences between IllustrisTNG and SIMBA in the appearance and statistics of HI because both simulations have been normalised to the same UVB strength as described in Appendix~\ref{sec:uv_correction}.  The cosmic HI distribution is robustly reproduced by many hydrodynamic simulations at high redshift \citep{theuns98,dave99,altay11,rahmati13}. Part of the visual difference owes to the jet-mode AGN feedback in SIMBA traveling many virial radii and heating voids, leading to much lower column densities \citep{sorini21,tillman_etal22}. AGN feedback is known to affect the HI column density statistics at low redshifts \citep{2017ApJ...835..175G,Burkhart_2022}. A detailed discussion of the differences between Simba and IllustrisTNG $z=0.1$ HI column density distributions, and how different implementations of AGN feedback alter the HI distributions, can be found in \citet{tillman_etal22}.  We further discuss the shape of the HI distribution in the Appendix, which is also different, as visible in Figure \ref{fig2}.  

\subsection {OVII Column Density Distributions }\label{3.2}

Figure~\ref{fig3} shows the CDDFs of OVII for sightlines with the same HI column densities as in K19, for both fiducial runs of the IllustrisTNG and SIMBA simulations. Both simulations show more OVII absorbers for sightlines with lower $N_\mathrm{HI}$. 
The $N_\mathrm{OVII}$ CDDF peaks at around $\log_{10} (N_{\rm{OVII}}/{\rm cm^{-2}}) \approx 12-14$, where sightlines with higher $\mathrm{N_{HI}}$ values peak at higher $N_{\rm OVII}$. Compared to the estimated $\log_{10} (N_\mathrm{OVII}/{\rm cm^{-2}}) \approx 15$ from K19, the peaks of the $N_\mathrm{OVII}$ values of the simulation are 1-2 orders of magnitude smaller. 

There are significant differences in $N_\mathrm{OVII}$ CDDF between IllustrisTNG and SIMBA. The peak in SIMBA occurs at lower $N_\mathrm{OVII}$, compared to IllustrisTNG. Thus, it is more unlikely to find OVII absorbers with the observed column density in the SIMBA runs. For the SIMBA runs, the distributions also drop sharply at $\log_{10} (N_\mathrm{OVII}/\mathrm{cm^{-2}}) \approx 15$, whereas for IllustrisTNG, this drop-off varies with $N_\mathrm{HI}$. This drop-off occurs at higher $N_\mathrm{OVII}$ for sightlines with higher $N_\mathrm{HI}$. 

We also show the CDDFs separately for OVII in filaments and halo outskirts in both simulations. For a given $N_\mathrm{HI}$ value, the halo outskirts tend to have slightly less OVII absorbers at low OVII column densities, but the trend is not significant. 

\begin{figure}
    \centering
    \includegraphics[width=0.48\textwidth]{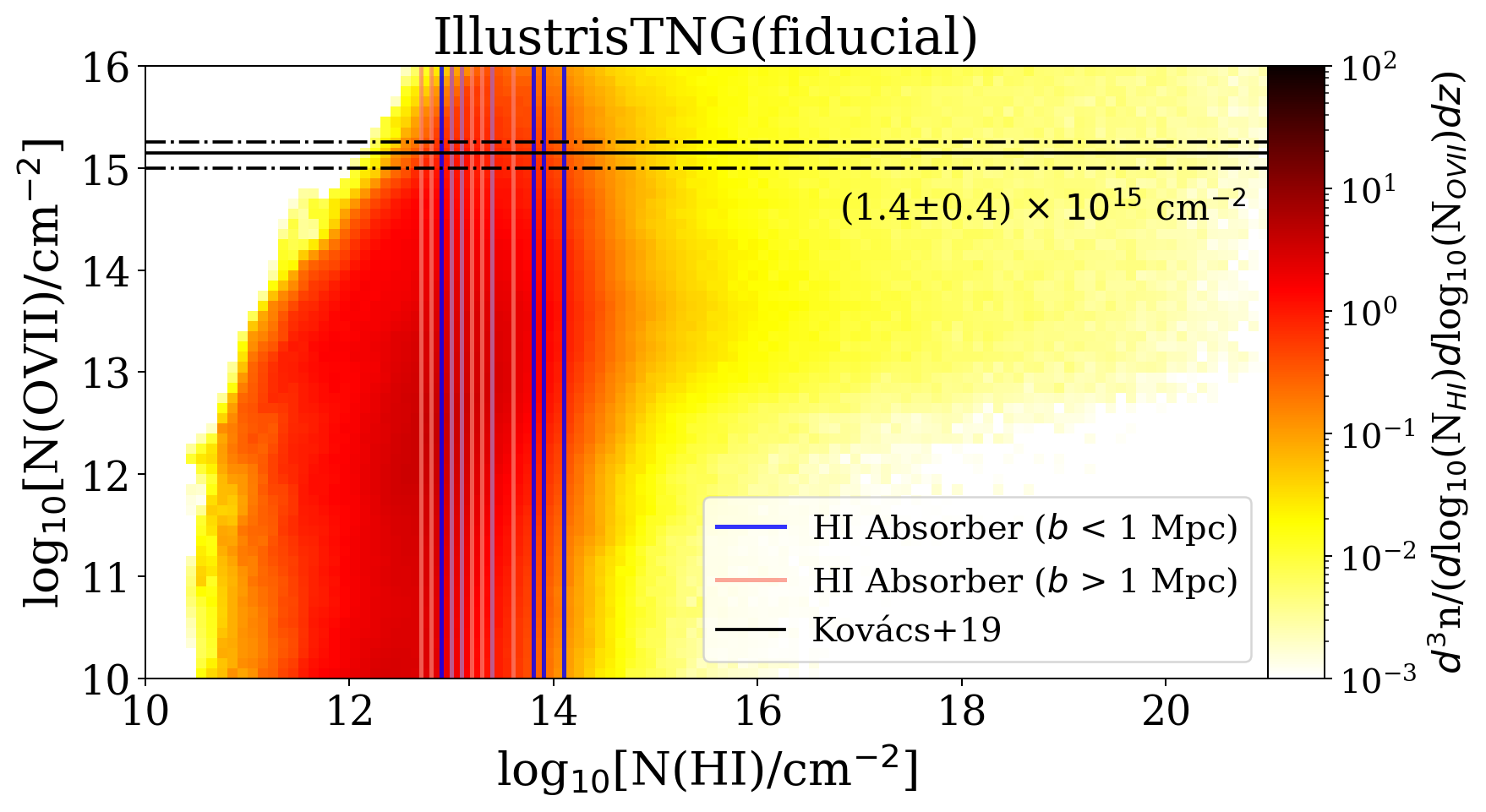}
    \includegraphics[width=0.48\textwidth]{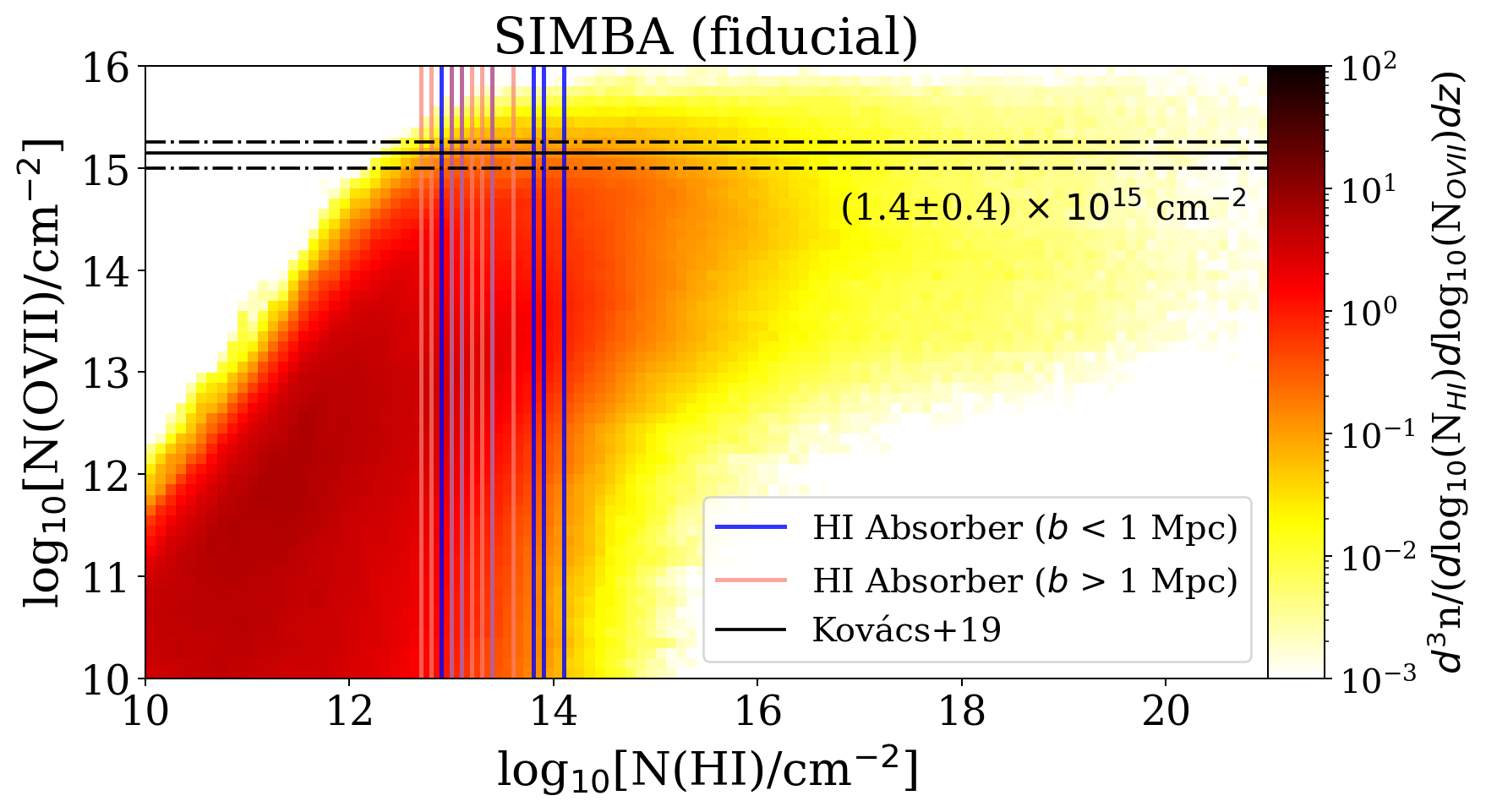}
    \caption{Two-dimensional CDDFs for the $\log_{10}(N_\mathrm{HI}/\mathrm{cm^{-2}})$ plotted against $\log_{10}(N_\mathrm{OVII}/\mathrm{cm^{-2}})$ for IllustrisTNG (top) and SIMBA (bottom) fiducial runs. We simultaneously overplot the expected OVII column density from the observation result of ($1.4 \pm 0.4$)$\times 10^{15}$ cm$^{-2}$ (black) in K19, as well as vertical lines corresponding to the column densities of the HI absorbers (pink for filaments, blue for halo outskirts) with a bin size of 0.1 dex. Note that the heat map indicates the column density distribution function for the pixels in each bin. We observe that in the lower regime of column density values especially, the distributions look quite different between the two simulations. This is a consequence of the different feedback physics implementations of the two simulations. We focus more closely on the differences in the regime closer to the observed K19 OVII column density.}
    \label{fig4}
\end{figure}

In Figure~\ref{fig4} we show the two-dimensional CDDF $d^3n/(d\log N_\mathrm{HI} d\log N_\mathrm{OVII} dz)$ for both IllustrisTNG and SIMBA. For a given $N_\mathrm{HI}$ value within the range of $\log_{10}(N_\mathrm{HI}/{\rm cm^{-2}}) \in [12.7, 14.1]$, there is a very weak positive correlation between $N_\mathrm{OVII}$ and $N_\mathrm{HI}$ with large scatter. The correlation is slightly stronger in the SIMBA run compared to the IllustrisTNG run. 

\begin{figure}
\centering
\includegraphics[width=0.48\textwidth]{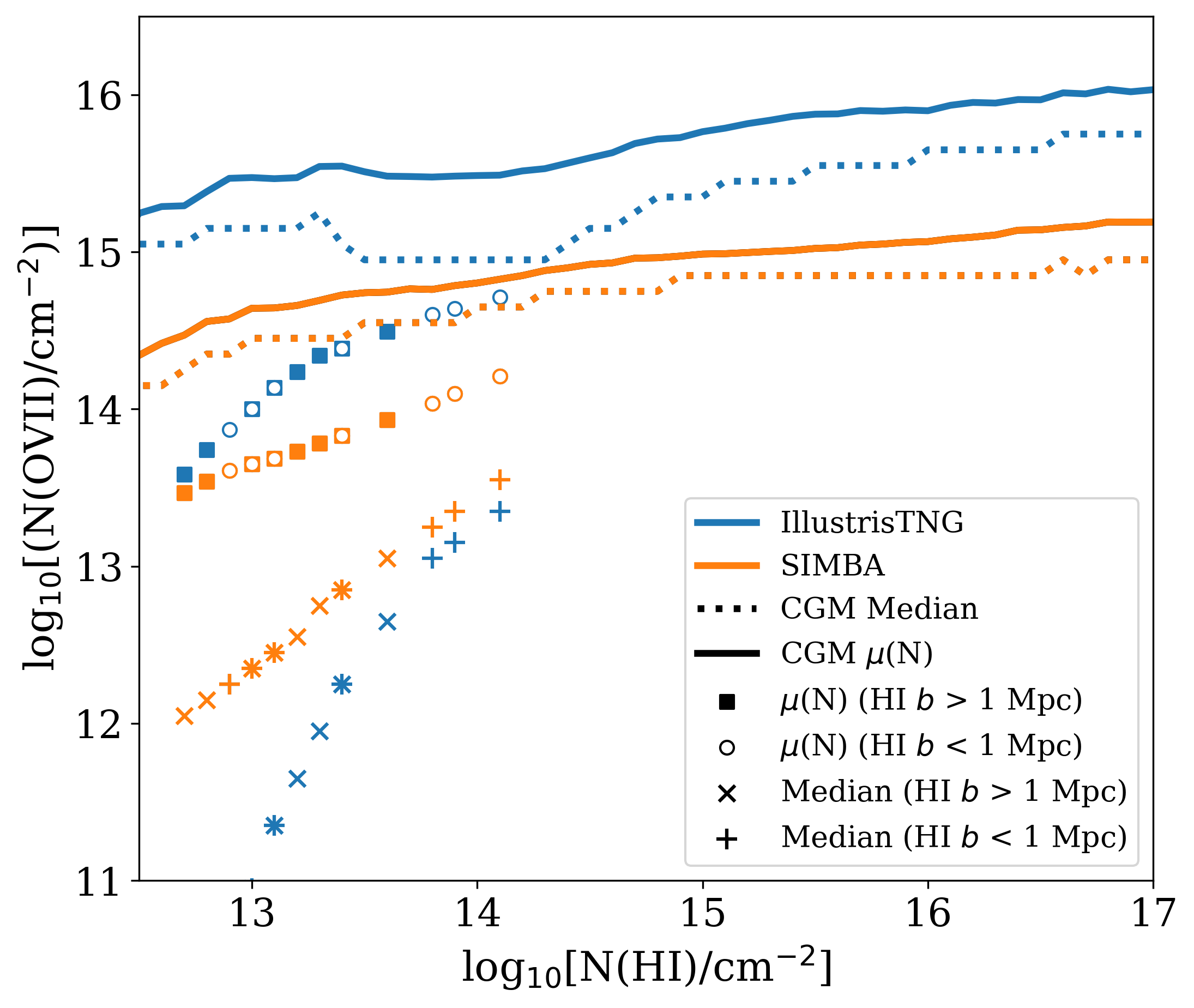}
\caption{Comparison of the mean $N_\mathrm{OVII}$ for different selection criteria for IllustrisTNG (blue) and SIMBA (orange) as a function of $N_\mathrm{HI}$. The solid lines indicate the mean $\mu$ of non-zero $N_\mathrm{OVII}$ distributions at corresponding incremental 0.1 dex $N_\mathrm{HI}$ values for 300 kpc CGM regions around galaxy halo locations. The dotted lines show the medians for these same 300 kpc distributions rounded to the nearest midpoint bin value. In the solid squares, we only present the average of the full OVII column density map at corresponding HI absorbers with an impact parameter $b > 1$ Mpc. In the empty circles, we show the average of the full OVII column density map at corresponding HI absorbers with an impact parameter $b < 1$ Mpc. There is overlap between these HI absorber separations. We show the median values of the full OVII column density maps at corresponding HI absorbers along the same division ($\times$ marker for $b > 1$ Mpc and $+$ marker for $b < 1$ Mpc). Additionally, we observe that the average $N_\mathrm{OVII}$ column densities for full maps fall much lower than the column densities for the 300 kpc CGM-only regions for both IllustrisTNG and SIMBA. That said, we see higher $\mu({N_\mathrm{OVII}})$ for IllustrisTNG than SIMBA throughout. 
}
\label{fig5}
\end{figure}

Figure~\ref{fig5} shows the mean of the $\log_{10} (N_\mathrm{OVII}/\mathrm{cm^{-2}})$ distribution at given $\log_{10} (N_\mathrm{HI}/\mathrm{cm^{-2}})$ values that correspond to the K19 absorbers. It shows that means of the $N_\mathrm{OVII}$ increase monotonically with $N_\mathrm{HI}$. In addition, we show separately the absorbers with impact parameter from their nearest galaxies $b > 1$~Mpc (solid squares) and $b>1$~Mpc (empty circles).  The $b = 1$~Mpc threshold is chosen to match with impact parameter binning in K19, which resulted in approximately the same number of absorbers in each bin (c.f. Sec~\ref{2.5}). Regardless of the impact parameter from the galaxies, the absorbers all follow the same monotonic trend. The IllustrisTNG values are consistently higher than the SIMBA values.  

Note that $b = 1$~Mpc is much larger than the typical virial radius of the galaxies in the K19 sample. The true CGM is expected to be much closer to the galaxies. In the same figure we also show mean and median of $N_\mathrm{OVII}$ as a function of $N_\mathrm{HI}$ within $b = 300$~kpc of the simulated galaxies. At the same $N_\mathrm{HI}$, the mean $N_\mathrm{OVII}$ values within $300$~kpc are consistently higher by an order of magnitude than the halo outskirts and WHIM absorbers, for both IllustrisTNG and SIMBA, suggesting that the $N_\mathrm{OVII}$ distribution at a given $N_\mathrm{HI}$ is highly non-gaussian, with a longer tail at the high end of $N_\mathrm{OVII}$. 

\subsection {Dependence of OVII Column Densities on Feedback Physics} \label{3.3}

\begin{figure*}
    \centering
    \includegraphics[width=0.76\textwidth]{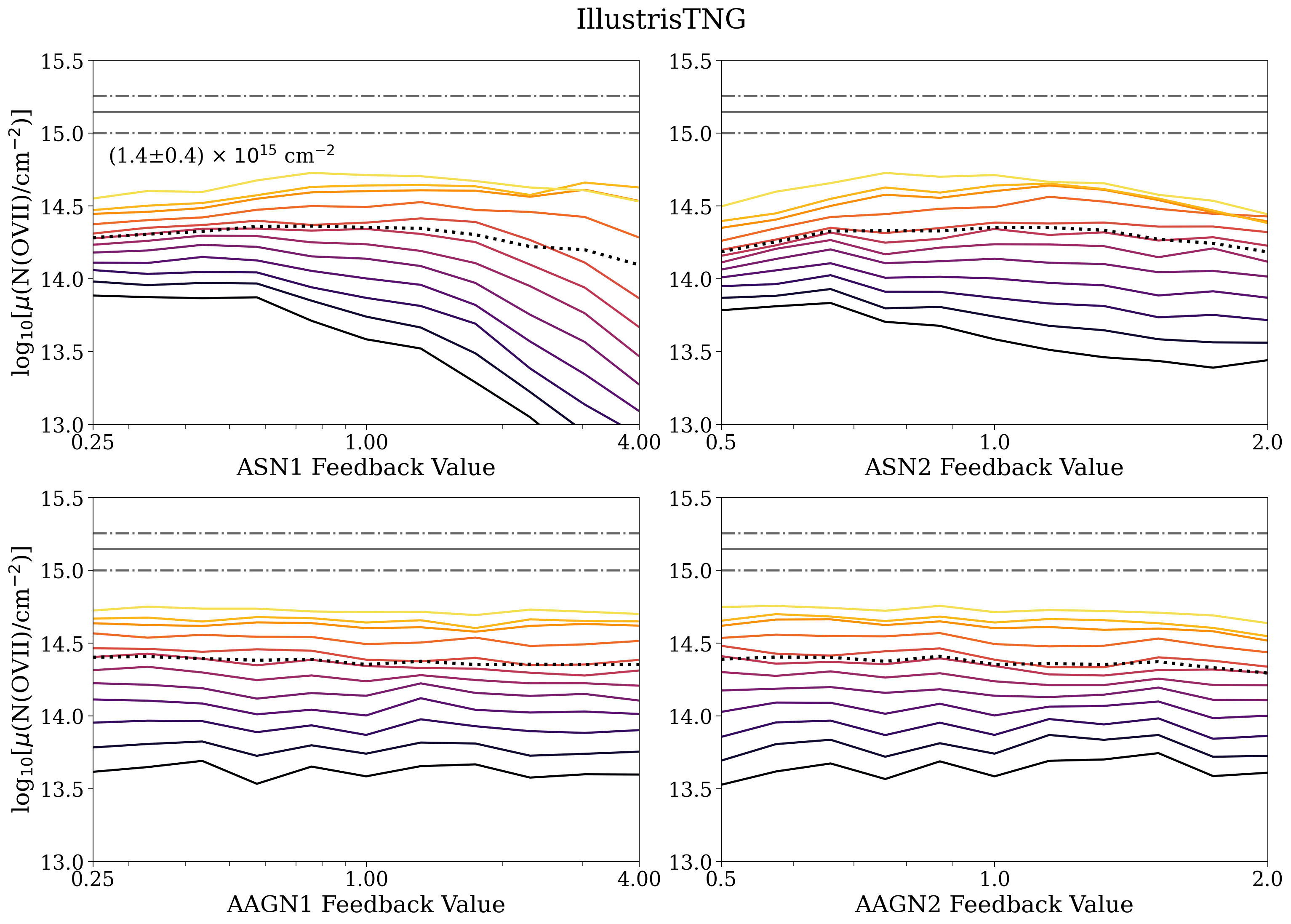}
    \includegraphics[width=0.76\textwidth]{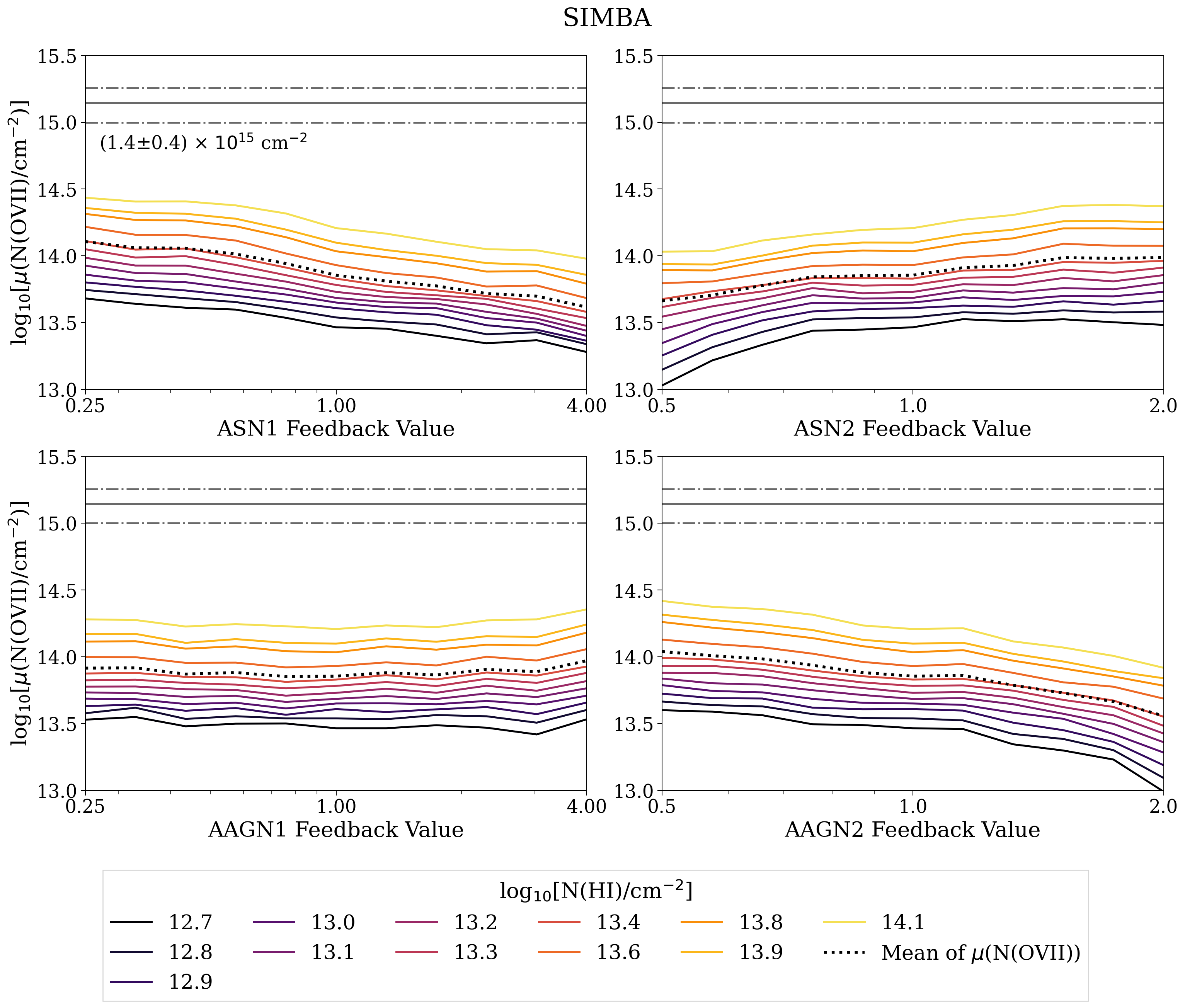}
    \caption{Dependence of Feedback on the OVII column densities measured at the values of observed HI column densities. The top 4 panels show the results for the IllustrisTNG runs, while the bottom 4 show the SIMBA results. In each panel, the horizontal solid and dot-dashed lines show respectively the mean and $1\sigma$ scatter of the observed $N_{\rm OVII}$ value from K19. The coloured lines show the means of the $N_\mathrm{OVII}$ distributions from the simulation for each observed $N_\mathrm{HI}$ value, with the colours indicating the relative level of $N_\mathrm{HI}$. The dotted line shows the mean of the simulated $N_\mathrm{OVII}$ estimates across all HI sightlines. 
    It shows that $N_\mathrm{OVII}$ is most dependent on varying SN1 and SN2 feedback in IllustrisTNG (SN energy per unit stellar mass and SN wind speed, respectively). More SN energy per stellar mass results in lower $N_\mathrm{OVII}$, especially for lower $N_\mathrm{HI}$ values. Varying AGN feedback in IllustrisTNG has virtually no impact on $N_\mathrm{OVII}$. For SIMBA, all feedback modes except AGN1 (momentum flux of AGN outflow) result in changes in $N_\mathrm{OVII}$, though the dependences are quite weak. For all ranges of feedback parameters explored, none produce high enough $N_\mathrm{OVII}$ to be consistent with observed value in K19.}
    \label{fig6}
\end{figure*}

Figure~\ref{fig6} shows the estimated OVII column densities at each observed HI sightline across all values of the CAMELS feedback parameters presented in Section~\ref{2.2}. We note that for all ranges of feedback parameters, the OVII column densities are still more than a magnitude below the observed estimates in K19, indicating that the differences we see between simulations and observations are unlikely due to feedback physics. Below, we examine in detail how the OVII column density depends on each CAMELS feedback mode. 

\subsubsection{Dependence on SN Feedback}\label{3.3.1}

The top left panel in Figure~\ref{fig6} shows the estimated OVII column densities at each observed HI sightline across all values of $A_{\mathrm{SN1}}$ feedback for both IllustrisTNG and SIMBA simulations. In IllustrisTNG, the $A_{\mathrm{SN1}}$ represents the amount of energy in the SN feedback per unit stellar mass, and in SIMBA, it represents the mass loading factor. The OVII column density in both IllustrisTNG and SIMBA depends on this $A_{\mathrm{SN1}}$ parameter. To show that, we plot the ``stacked'', or mean of means estimate of $\log_{10} (N_\mathrm{OVII}/\mathrm{cm^{-2}})$ at the observed $N_\mathrm{HI}$ values. Increasing $A_{\mathrm{SN1}}$ feedback decreases the estimated OVII column densities for all HI sightlines in the SIMBA runs, and for HI sightline column densities $\log_{10}(N_\mathrm{HI}/\mathrm{cm^{-2}}) \lesssim 13.6$ for the IllustrisTNG run. We interpret this trend as stronger suppression of star formation with increasing SN energy output, leading to lower oxygen yields and OVII. This effect is smaller at higher $N_\mathrm{HI}$ in more massive halos in IllustrisTNG, where SN feedback is less effective in quenching star formation.

We also examine the dependence on $A_{\mathrm{SN2}}$, representing the wind speed of SN feedback for both IllustrisTNG and SIMBA. Again, the ``stacked'' value of OVII column density is lower than the value in K19 in all IllustrisTNG and SIMBA runs. In IllustrisTNG, we observe a slight decrease in OVII for $\log_{10}(N_\mathrm{HI}/\mathrm{cm^{-2}}) \lesssim 13.6$, and a decrease in the remaining sightlines starting at $A_{\mathrm{SN2}} \approx 1$. In SIMBA, we see an increase in OVII as feedback increases across all sightlines, though this trend is not strictly monotonic, especially for HI absorbers with low column densities. In IllustrisTNG, the HI absorber sightlines that are most affected by the feedback fall into the regime of the filament rather than the halo outskirt, indicating that the halo outskirt may be more resilient to varying SN wind speed and mass loading factor. In SIMBA, both the halo outskirt and filament regimes appear to be affected by wind speed and wind energy flux modulation, though not uniformly.

\subsubsection{Dependence on AGN Feedback}\label{3.3.2}
The $A_{\mathrm{AGN1}}$ feedback represents the energy of the AGN kinetic feedback in the IllustrisTNG runs and the momentum flux of the AGN outflow in the SIMBA runs. 
For $A_{\mathrm{AGN1}}$ feedback, we observe negligible effects of varied feedback for both simulation suites. For IllustrisTNG, there are no clear trends, suggesting that the kinetic feedback power adjustment has a limited impact on the OVII and HI column density distributions. For SIMBA, there is a small increase for the most extreme feedback values $A_{\mathrm{AGN1}} \gtrsim 3$, but otherwise, the effect of quasar and jet outflows is negligible.

The $A_{\mathrm{AGN2}}$ feedback represents the temperature of the gas heated by each AGN outburst for IllustrisTNG and the speed of the jet outflow in the SIMBA runs. Because of the difference in the physical meaning of $A_{\mathrm{AGN2}}$ between the two suites, we see significant differences in the $N_\mathrm{OVII}$ distributions between the two. For IllustrisTNG, there is a very slight decrease in $N_\mathrm{OVII}$ as $A_{\mathrm{AGN2}}$ increases, especially at higher HI absorber sightline values ($\log_{10}(N_\mathrm{HI}/\mathrm{cm^{-2}}) \gtrsim 13.1$) and for stronger feedbacks ($A_{\mathrm{AGN1}} \gtrsim 1$). For SIMBA, $A_{\mathrm{AGN2}}$ more strongly decreases OVII for all sightlines as feedback increases. We interpret this result as the OVII being more responsive to the speed of continuously driven AGN jets, where faster jets quench more star formation and thus inhibit OVII formation. 

\begin{figure}
    \centering
    \includegraphics[width=0.5\textwidth]{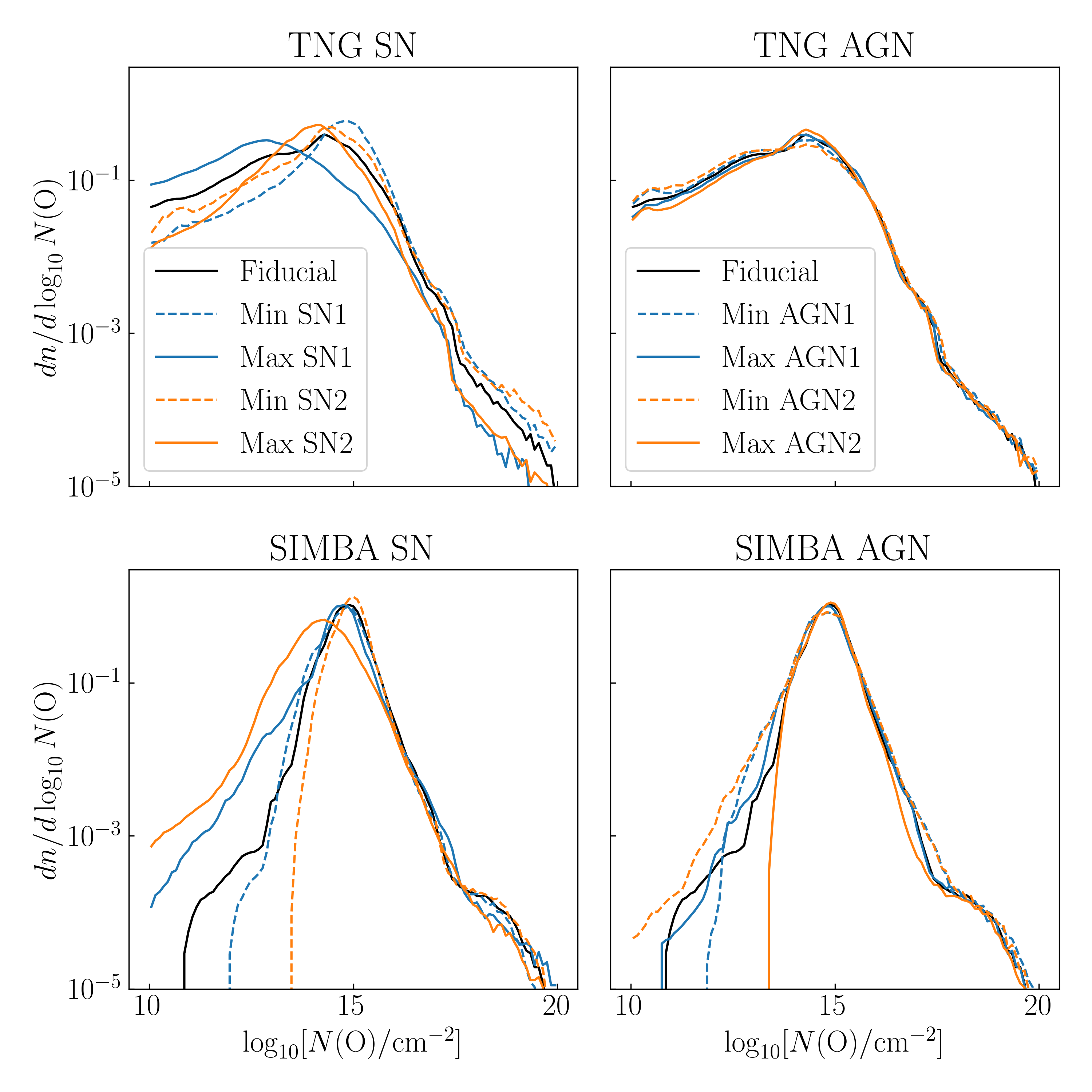}
    \caption{Distributions of oxygen column densities $N_\mathrm{O}$ for the IllustrisTNG (top panels) and SIMBA (bottom panels) runs. The left panels show the distributions for min and max SN1 and SN2 runs. The right panels show the distributions for min and max AGN1 and AGN2 runs. The figure shows that for the IllustrisTNG run, the oxygen column density distributions are most sensitive to varying the $A_\mathrm{SN1}$ and $A_\mathrm{SN2}$ parameters, which represent the SN energy per unit stellar mass and the SN wind speed, respectively. The $N_\mathrm{O}$ distribution is insensitive to AGN feedback. On the other hand, for the SIMBA runs, all SN and AGN feedback modes have a considerable impact on the $N_\mathrm{O}$ distribution at low column densities (for $\log_{10}(N_\mathrm{O}/\mathrm{cm^{-2}})  < 15 $). The dependence on SN and AGN feedback in $N_\mathrm{O}$ for both IllustrisTNG and SIMBA is similar to that of $N_\mathrm{OVII}$, suggesting that feedback does very little in changing the ionisation state of oxygen. The dependence on feedback in $N_\mathrm{OVII}$ is likely due to the suppression of star formation and hence total oxygen production.} 
    \label{fig:O_hist}
\end{figure}

\begin{figure}
    \centering
    \includegraphics[width=0.5\textwidth]{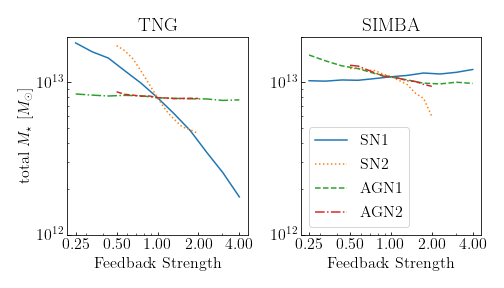}
    \caption{We show the total stellar mass at $z=0.154$ produced in the CAMEL simulation box and its dependence on the feedback strengths, for both the IllustrisTNG runs (left panel) and the SIMBA runs (right panel).   For TNG, the increasing $A_{\rm SN1}$ (blue line) and $A_{\rm SN2}$ (orange dotted line) decreases the total stellar mass, while increasing $A_{\rm AGN1}$ (green dashed ine) and $A_{\rm AGN2}$ (red dashed-dotted line) results in almost no change in stellar mass. On the contrary, for the SIMBA runs, increasing $A_{\rm SN1}$ leads to slight increase in stellar mass, while increasing the other feedback parameters leads to lower stellar mass.  } 
    \label{fig:mstar}
\end{figure}

\subsubsection{Origin of the Dependence on Feedback}\label{sec:3.3}

The different dependence of $N_\mathrm{OVII}$ mainly comes through the dependence of star formation quenching on feedback implementations. In Figure~\ref{fig:O_hist}, we show the distributions of the column densities of {\em all} oxygen species for the extreme runs of the 4 feedback modes for both  IllustrisTNG and SIMBA.
For the IllustrisTNG run, the oxygen column density distributions are sensitive to SN1 and SN2 feedback, representing the SN energy per unit stellar mass and the SN wind speed, respectively, while insensitive to AGN feedback. For the SIMBA runs, all SN and AGN feedback modes have considerable impact on the $N_\mathrm{O}$ distribution at low column densities (for $\log_{10}(N_\mathrm{O}/\mathrm{cm^{-2}})  < 15 $, but minimal impact otherwise. 

The column densities of total oxygen O show the same qualitative dependence on feedback as that for OVII. This can be explained by the reduction in the production of metals in runs with stronger SN feedback, as the star formation and thus metal production is quenched more strongly, reducing the total amount of oxygen and its ionised species. In Figure~\ref{fig:mstar}, we show the dependence of total stellar mass produced in the box on the strengths of the 4 feedback modes. For IllustrisTNG, the stellar mass is most sensitive to SN1 and SN2 feedback, but almost insensitive to AGN1 and AGN2 feedback. For SIMBA, increasing SN1 and SN2 feedback leads to mildly increasing and decreasing stellar mass, respectively; increasing both AGN1 and AGN2 feedback leads to a modest drop in stellar mass. These trends are consistent with the feedback dependence of $N_{\mathrm{O}}$ in Figure~\ref{fig:O_hist}. 
The similar dependence of feedback in O, OVII (and stellar mass) implies that feedback has a limited impact on the ionisation state of oxygen in WHIM. 

Note that the O column density distribution also behaves quite differently between the IllustrisTNG and the SIMBA runs. Firstly, the IllustrisTNG distribution has lower column density absorbers than SIMBA, for all feedback runs. Secondly, the column density distributions are more sensitive to SN feedback in the SIMBA runs than in the IllustrisTNG runs, especially at the lower column densities. This highlights the differences in the sub-grid modeling of feedback between IllustrisTNG and SIMBA and their predictions on WHIM properties. 

\section{Conclusions and Discussion}\label{summary}

Using the CAMEL simulation suite, we study the OVII primarily arising from the Warm Hot Intergalactic Medium (WHIM), how this ion depends on feedback by supernovae (SN) and 
active galactic nuclei (AGN), and whether these simulations can reproduce the observed OVII X-ray absorption signal detected along the H1821+643 quasar sight line obtained by stacking known HI absorbers \citep[][K19]{kovacs19}.  Here are our main findings: 
\begin{enumerate}

    \item 
    For all ranges of SN and AGN feedback parameters in CAMELS, the $N_\mathrm{{OVII}}$ values for the WHIM are 1-2 orders of magnitude below the {\em Chandra} observation of  $N_\mathrm{{OVII}}$ in \citep[][K19]{kovacs19}. In particular, the maximum value of $N_\mathrm{{OVII}}$ is below $N_\mathrm{{OVII}}$ of $(1.4 \pm 0.4) \times 10^{15}\,\mathrm{cm^{-2}}$, the observed value (see Figure~\ref{fig6}).
    
    \item 
    The OVII column density is most sensitive to the energy of the SN feedback per unit stellar mass in the IllustrisTNG runs, and the SN mass loading factor in SIMBA.   Other modes of SN feedback and most AGN feedback explored in the CAMEL simualtion suite do little to affect OVII column density. 
    
    \item The OVII column density is insensitive to the AGN feedback energy in both IllustrisTNG and SIMBA, which can be attributed to the relative spatial rarity of AGN relative to the stellar sources of SN feedback. On the other hand, increasing the AGN jet speed in the SIMBA runs lowers OVII, indicating these jets can impact a significant cosmic volume. 
    
    \item The relationship of OVII-HI column densities follows a similar relationship if found inside or outside a 1~Mpc radius from galaxies, in agreement with K19.  This owes to the OVII arising almost entirely from the WHIM.  In contrast, the CGM within 300 kpc of galaxies shows significantly stronger OVII for a given HI absorption strength, but this is not a dominant contribution to the H1821+643 sightline (see Figures \ref{fig1} and \ref{fig5}).
    

\end{enumerate}


The K19 estimate on the OVII column densities inferred from stacking of X-ray absorption lines is higher than nearly all of the sightlines in the CAMEL simulations with varying feedback physics for similar values of HI column densities as in the observation. In addition, in the simulation, not any one slightline dominates the average OVII column density estimate. In Appendix~\ref{sec:convergence}, we show that the differences between simulations and observations are unlikely due to cosmic variance since going to a large simulation box will only explain $0.2$ dex differences in the OVII column densities. 

This suggests a tension between the observed OVII column density estimate in K19 and predictions from our simulations. This can mean either (1) the gas physics models adopted in both IllustrisTNG and SIMBA are insufficient to produce enough oxygen column densities, or (2) the K19 observed column density is dominated by a few strong absorbers that skew the stacked measurement towards high column densities, since individual absorbers were not detected. Exploring other astrophysical processes not varied in CAMELS,  such as turbulent mixing, or stellar yield, may help distinguish between these two scenarios. Additionally, studying more sightlines will also help to resolve whether the H1821+643 sightline is brighter than other sightlines. To this end,  we analyze additional sightlines based on {\em Chandra} archival observations in a follow-up study and compare the OVII column densities and upper limits with those observed in \citet{kovacs19}. The detailed comparison between the observed sightlines will be presented in an upcoming study.  

We see that supernovae feedback typically affects the OVII distributions more than AGN feedback. This is because (1) supernovae feedback sources are more widely distributed than AGN sources, which are typically centered on rare, massive halos, and (2) supernovae are more effective in suppressing star formation in the more abundant, lower mass halos. Lower $N_\mathrm{{OVII}}$ corresponding to lower $N_\mathrm{{HI}}$ sightlines are more affected by the $A_{\mathrm{SN1}}$ feedback since these $N_\mathrm{{OVII}}$ values reside in the more sparse filamentary structures, rather than the high $N_\mathrm{{OVII}}$, which are clustered around massive halos. A larger simulation box with more massive, group-size halos will allow us better to assess the impact of AGN feedback on WHIM. 

The differences in the column density distributions between IllustrisTNG and SIMBA within CAMELS highlight that the feedback physics implementations have not yet converged. Both produce quite different OVII column density distributions despite having similar strengths of the feedback parameter. Specifically, the SIMBA runs generate consistently lower numbers of OVII absorbers at high column densities than IllustrisTNG. This suggests that the jet feedback in SIMBA runs (compared to kinetic feedback in IllustrisTNG) is more efficient in suppressing the production of stars and oxygen in more massive halos. This also suggests that the relationship between OVII and HI column densities in the WHIM are sensitive to, and thus can be used to constrain feedback physics. 

Ongoing analysis of {\em Chandra} observation of similar systems, but with deeper exposure, will help in resolving the differences in OVII column densities between simulation and observation. 
Future X-ray instruments with high spectral resolution and sensitivity, 
such as 
an ARCUS-like \citep {arcus} mission dedicated to X-ray absorption,
the \textit{Athena X-Ray Observatory} \citep{nandra2013hot}, 
and the proposed \textit{Line Emission Mapper} (LEM)\footnote{\url{https://www.cfa.harvard.edu/spaces/lem/}}, can provide more accurate measurements of absorption line spectra (for WHIM emission, see \citealt{parimbelli_etal22} for CAMELS prediction for \textit{Athena}), all necessary for uncovering a more realistic picture of the complex gas structures present in the universe and the locations of elusive low-redshift baryons. 

\section*{Acknowledgements}
We thank the anonymous referee whose comments improve the presentation of the paper. We also Nastasha Wijers, Joop Schaye, and Johnny Dorigo Jones for their commments. We gratefully acknowledge the CAMELS team for publicly releasing CAMELS data. We also acknowledge the Center for Computational Astrophysics at the Flatiron Institute for providing the computing facilities for making the CAMELS maps and the Yale Center for Research Computing for enabling our use of High-Performance Computing. This work is supported by NSF grant AST-2206055. \'A.B. acknowledges support from the Smithsonian Institution and the Chandra High Resolution Camera Project through NASA contract NAS8-03060. 
O.E.K. is supported by the GACR grant 21-13491X. 

\section*{Data Availability}
The instructions for accessing the CAMELS data \citep{camelsrelease} can be found here: \url{https://camels.readthedocs.io/en/latest/data_access.html}. Accessing the Trident code suite \citep{hummels2017} to enable manipulation and analysis of CAMELS data can be found here: \url{https://trident.readthedocs.io/en/latest/}.



\bibliographystyle{mnras}
\bibliography{references} 

\begin{thebibliography}{}
\makeatletter
\relax
\def\mn@urlcharsother{\let\do\@makeother \do\$\do\&\do\#\do\^\do\_\do\%\do\~}
\def\mn@doi{\begingroup\mn@urlcharsother \@ifnextchar [ {\mn@doi@}
  {\mn@doi@[]}}
\def\mn@doi@[#1]#2{\def\@tempa{#1}\ifx\@tempa\@empty \href
  {http://dx.doi.org/#2} {doi:#2}\else \href {http://dx.doi.org/#2} {#1}\fi
  \endgroup}
\def\mn@eprint#1#2{\mn@eprint@#1:#2::\@nil}
\def\mn@eprint@arXiv#1{\href {http://arxiv.org/abs/#1} {{\tt arXiv:#1}}}
\def\mn@eprint@dblp#1{\href {http://dblp.uni-trier.de/rec/bibtex/#1.xml}
  {dblp:#1}}
\def\mn@eprint@#1:#2:#3:#4\@nil{\def\@tempa {#1}\def\@tempb {#2}\def\@tempc
  {#3}\ifx \@tempc \@empty \let \@tempc \@tempb \let \@tempb \@tempa \fi \ifx
  \@tempb \@empty \def\@tempb {arXiv}\fi \@ifundefined
  {mn@eprint@\@tempb}{\@tempb:\@tempc}{\expandafter \expandafter \csname
  mn@eprint@\@tempb\endcsname \expandafter{\@tempc}}}

\bibitem[\protect\citeauthoryear{{Altay}, {Theuns}, {Schaye}, {Crighton}  \&
  {Dalla Vecchia}}{{Altay} et~al.}{2011}]{altay11}
{Altay} G.,  {Theuns} T.,  {Schaye} J.,  {Crighton} N. H.~M.,   {Dalla Vecchia}
  C.,  2011, \mn@doi [\apjl] {10.1088/2041-8205/737/2/L37}, \href
  {https://ui.adsabs.harvard.edu/abs/2011ApJ...737L..37A} {737, L37}

\bibitem[\protect\citeauthoryear{{Alvarez}, {Randall}, {Bourdin}, {Jones}  \&
  {Holley-Bockelmann}}{{Alvarez} et~al.}{2018}]{alvarez18}
{Alvarez} G.~E.,  {Randall} S.~W.,  {Bourdin} H.,  {Jones} C.,
  {Holley-Bockelmann} K.,  2018, \mn@doi [\apj] {10.3847/1538-4357/aabad0},
  \href {https://ui.adsabs.harvard.edu/abs/2018ApJ...858...44A} {858, 44}

\bibitem[\protect\citeauthoryear{{Bahcall} et~al.,}{{Bahcall}
  et~al.}{1993}]{bahcall93}
{Bahcall} J.~N.,  et~al., 1993, \mn@doi [\apjs] {10.1086/191797}, \href
  {https://ui.adsabs.harvard.edu/abs/1993ApJS...87....1B} {87, 1}

\bibitem[\protect\citeauthoryear{{Bird}}{{Bird}}{2017}]{Bird:2017}
{Bird} S.,  2017, {FSFE: Fake Spectra Flux Extractor}, Astrophysics Source Code
  Library, record ascl:1710.012 (\mn@eprint {ascl} {1710.012})

\bibitem[\protect\citeauthoryear{{Bird}, {Haehnelt}, {Neeleman}, {Genel},
  {Vogelsberger}  \& {Hernquist}}{{Bird} et~al.}{2015}]{Bird:2015}
{Bird} S.,  {Haehnelt} M.,  {Neeleman} M.,  {Genel} S.,  {Vogelsberger} M.,
  {Hernquist} L.,  2015, \mn@doi [\mnras] {10.1093/mnras/stu2542}, \href
  {https://ui.adsabs.harvard.edu/abs/2015MNRAS.447.1834B} {447, 1834}

\bibitem[\protect\citeauthoryear{{Bregman}}{{Bregman}}{2007}]{bregman07}
{Bregman} J.~N.,  2007, \mn@doi [\araa]
  {10.1146/annurev.astro.45.051806.110619}, \href
  {http://adsabs.harvard.edu/abs/2007ARA%26A..45..221B} {45, 221}

\bibitem[\protect\citeauthoryear{Burkhart, Tillman, Gurvich, Bird, Tonnesen,
  Bryan, Hernquist  \& Somerville}{Burkhart et~al.}{2022}]{Burkhart_2022}
Burkhart B.,  Tillman M.,  Gurvich A.~B.,  Bird S.,  Tonnesen S.,  Bryan G.~L.,
   Hernquist L.,   Somerville R.~S.,  2022, \mn@doi [The Astrophysical Journal
  Letters] {10.3847/2041-8213/ac7e49}, 933, L46

\bibitem[\protect\citeauthoryear{{Burles}, {Nollett}  \& {Turner}}{{Burles}
  et~al.}{2001}]{burles01}
{Burles} S.,  {Nollett} K.~M.,   {Turner} M.~S.,  2001, \mn@doi [\apjl]
  {10.1086/320251}, \href
  {https://ui.adsabs.harvard.edu/abs/2001ApJ...552L...1B} {552, L1}

\bibitem[\protect\citeauthoryear{{Cooke}, {Pettini}, {Jorgenson}, {Murphy}  \&
  {Steidel}}{{Cooke} et~al.}{2014}]{cooke2014}
{Cooke} R.~J.,  {Pettini} M.,  {Jorgenson} R.~A.,  {Murphy} M.~T.,   {Steidel}
  C.~C.,  2014, \mn@doi [\apj] {10.1088/0004-637X/781/1/31}, \href
  {https://ui.adsabs.harvard.edu/abs/2014ApJ...781...31C} {781, 31}

\bibitem[\protect\citeauthoryear{{Copi}, {Schramm}  \& {Turner}}{{Copi}
  et~al.}{1995}]{copi95}
{Copi} C.~J.,  {Schramm} D.~N.,   {Turner} M.~S.,  1995, \mn@doi [Science]
  {10.1126/science.7809624}, \href
  {https://ui.adsabs.harvard.edu/abs/1995Sci...267..192C} {267, 192}

\bibitem[\protect\citeauthoryear{{Crain} et~al.,}{{Crain}
  et~al.}{2015}]{crain15}
{Crain} R.~A.,  et~al., 2015, \mn@doi [\mnras] {10.1093/mnras/stv725}, \href
  {http://adsabs.harvard.edu/abs/2015MNRAS.450.1937C} {450, 1937}

\bibitem[\protect\citeauthoryear{{Danforth} et~al.,}{{Danforth}
  et~al.}{2016}]{danforth16}
{Danforth} C.~W.,  et~al., 2016, \mn@doi [\apj] {10.3847/0004-637X/817/2/111},
  \href {https://ui.adsabs.harvard.edu/abs/2016ApJ...817..111D} {817, 111}

\bibitem[\protect\citeauthoryear{{Dav{\'e}}, {Hernquist}, {Katz}  \&
  {Weinberg}}{{Dav{\'e}} et~al.}{1999}]{dave99}
{Dav{\'e}} R.,  {Hernquist} L.,  {Katz} N.,   {Weinberg} D.~H.,  1999, \mn@doi
  [\apj] {10.1086/306722}, \href
  {https://ui.adsabs.harvard.edu/abs/1999ApJ...511..521D} {511, 521}

\bibitem[\protect\citeauthoryear{{Dav{\'e}}, {Angl{\'e}s-Alc{\'a}zar},
  {Narayanan}, {Li}, {Rafieferantsoa}  \& {Appleby}}{{Dav{\'e}}
  et~al.}{2019}]{dave19}
{Dav{\'e}} R.,  {Angl{\'e}s-Alc{\'a}zar} D.,  {Narayanan} D.,  {Li} Q.,
  {Rafieferantsoa} M.~H.,   {Appleby} S.,  2019, \mn@doi [\mnras]
  {10.1093/mnras/stz937}, \href
  {https://ui.adsabs.harvard.edu/abs/2019MNRAS.486.2827D} {486, 2827}

\bibitem[\protect\citeauthoryear{{Dorigo Jones} et~al.,}{{Dorigo Jones}
  et~al.}{2022}]{dorigo_jones_etal22}
{Dorigo Jones} J.,  et~al., 2022, \mn@doi [\mnras] {10.1093/mnras/stab3331},
  \href {https://ui.adsabs.harvard.edu/abs/2022MNRAS.509.4330D} {509, 4330}

\bibitem[\protect\citeauthoryear{{Eckert} et~al.,}{{Eckert}
  et~al.}{2015}]{eckert15}
{Eckert} D.,  et~al., 2015, \mn@doi [\nat] {10.1038/nature16058}, \href
  {https://ui.adsabs.harvard.edu/abs/2015Natur.528..105E} {528, 105}

\bibitem[\protect\citeauthoryear{{Fukugita}, {Hogan}  \& {Peebles}}{{Fukugita}
  et~al.}{1998}]{fukugita_etal98}
{Fukugita} M.,  {Hogan} C.~J.,   {Peebles} P.~J.~E.,  1998, \mn@doi [\apj]
  {10.1086/306025}, \href
  {https://ui.adsabs.harvard.edu/abs/1998ApJ...503..518F} {503, 518}

\bibitem[\protect\citeauthoryear{{Geller} \& {Huchra}}{{Geller} \&
  {Huchra}}{1989}]{geller_huchra89}
{Geller} M.~J.,  {Huchra} J.~P.,  1989, \mn@doi [Science]
  {10.1126/science.246.4932.897}, \href
  {https://ui.adsabs.harvard.edu/abs/1989Sci...246..897G} {246, 897}

\bibitem[\protect\citeauthoryear{{Gurvich}, {Burkhart}  \& {Bird}}{{Gurvich}
  et~al.}{2017}]{2017ApJ...835..175G}
{Gurvich} A.,  {Burkhart} B.,   {Bird} S.,  2017, \mn@doi [\apj]
  {10.3847/1538-4357/835/2/175}, \href
  {https://ui.adsabs.harvard.edu/abs/2017ApJ...835..175G} {835, 175}

\bibitem[\protect\citeauthoryear{{Hummels}, {Smith}  \& {Silvia}}{{Hummels}
  et~al.}{2017}]{hummels2017}
{Hummels} C.~B.,  {Smith} B.~D.,   {Silvia} D.~W.,  2017, \mn@doi [\apj]
  {10.3847/1538-4357/aa7e2d}, \href
  {https://ui.adsabs.harvard.edu/abs/2017ApJ...847...59H} {847, 59}

\bibitem[\protect\citeauthoryear{{Jannuzi} et~al.,}{{Jannuzi}
  et~al.}{1998}]{jannuzi98}
{Jannuzi} B.~T.,  et~al., 1998, \mn@doi [\apjs] {10.1086/313130}, \href
  {https://ui.adsabs.harvard.edu/abs/1998ApJS..118....1J} {118, 1}

\bibitem[\protect\citeauthoryear{{Johnson} et~al.,}{{Johnson}
  et~al.}{2019}]{johnson19}
{Johnson} S.~D.,  et~al., 2019, \mn@doi [\apjl] {10.3847/2041-8213/ab479a},
  \href {https://ui.adsabs.harvard.edu/abs/2019ApJ...884L..31J} {884, L31}

\bibitem[\protect\citeauthoryear{{Kaastra}, {Werner}, {Herder}, {Paerels}, {de
  Plaa}, {Rasmussen}  \& {de Vries}}{{Kaastra} et~al.}{2006}]{kaastra06}
{Kaastra} J.~S.,  {Werner} N.,  {Herder} J.~W.~A.~d.,  {Paerels} F.~B.~S.,  {de
  Plaa} J.,  {Rasmussen} A.~P.,   {de Vries} C.~P.,  2006, \mn@doi [\apj]
  {10.1086/507835}, \href
  {https://ui.adsabs.harvard.edu/abs/2006ApJ...652..189K} {652, 189}

\bibitem[\protect\citeauthoryear{{Kov{\'a}cs}, {Bogd{\'a}n}, {Smith}, {Kraft}
  \& {Forman}}{{Kov{\'a}cs} et~al.}{2019}]{kovacs19}
{Kov{\'a}cs} O.~E.,  {Bogd{\'a}n} {\'A}.,  {Smith} R.~K.,  {Kraft} R.~P.,
  {Forman} W.~R.,  2019, \mn@doi [\apj] {10.3847/1538-4357/aaef78}, \href
  {https://ui.adsabs.harvard.edu/abs/2019ApJ...872...83K} {872, 83}

\bibitem[\protect\citeauthoryear{{Lehner}, {Savage}, {Richter}, {Sembach},
  {Tripp}  \& {Wakker}}{{Lehner} et~al.}{2007}]{lehner07}
{Lehner} N.,  {Savage} B.~D.,  {Richter} P.,  {Sembach} K.~R.,  {Tripp} T.~M.,
   {Wakker} B.~P.,  2007, \mn@doi [\apj] {10.1086/511749}, \href
  {https://ui.adsabs.harvard.edu/abs/2007ApJ...658..680L} {658, 680}

\bibitem[\protect\citeauthoryear{{Mathur}, {Weinberg}  \& {Chen}}{{Mathur}
  et~al.}{2003}]{mathur03}
{Mathur} S.,  {Weinberg} D.~H.,   {Chen} X.,  2003, \mn@doi [\apj]
  {10.1086/344509}, \href
  {https://ui.adsabs.harvard.edu/abs/2003ApJ...582...82M} {582, 82}

\bibitem[\protect\citeauthoryear{Nandra, Barret, Barcons, Fabian  \& den
  Herder~et al.}{Nandra et~al.}{2013}]{nandra2013hot}
Nandra K.,  Barret D.,  Barcons X.,  Fabian A.,   den Herder~et al. J.-W.,
  2013, {The} {Hot} {and} {Energetic} {Universe:} {A} {White} {Paper}
  {presenting the science theme motivating the} {Athena+} {mission} (\mn@eprint
  {arXiv} {1306.2307})

\bibitem[\protect\citeauthoryear{{Nelson} et~al.,}{{Nelson}
  et~al.}{2018}]{nelson18}
{Nelson} D.,  et~al., 2018, \mn@doi [\mnras] {10.1093/mnras/sty656}, \href
  {https://ui.adsabs.harvard.edu/abs/2018MNRAS.477..450N} {477, 450}

\bibitem[\protect\citeauthoryear{{Nelson} et~al.,}{{Nelson}
  et~al.}{2019}]{nelson19}
{Nelson} D.,  et~al., 2019, \mn@doi [Computational Astrophysics and Cosmology]
  {10.1186/s40668-019-0028-x}, \href
  {https://ui.adsabs.harvard.edu/abs/2019ComAC...6....2N} {6, 2}

\bibitem[\protect\citeauthoryear{{Nicastro} et~al.,}{{Nicastro}
  et~al.}{2005}]{nicastro05}
{Nicastro} F.,  et~al., 2005, \mn@doi [\nat] {10.1038/nature03245}, \href
  {https://ui.adsabs.harvard.edu/abs/2005Natur.433..495N} {433, 495}

\bibitem[\protect\citeauthoryear{{Nicastro} et~al.,}{{Nicastro}
  et~al.}{2018}]{nicastro18}
{Nicastro} F.,  et~al., 2018, \mn@doi [\nat] {10.1038/s41586-018-0204-1}, \href
  {https://ui.adsabs.harvard.edu/abs/2018Natur.558..406N} {558, 406}

\bibitem[\protect\citeauthoryear{{Oppenheimer}, {Dav{\'e}}, {Katz}, {Kollmeier}
   \& {Weinberg}}{{Oppenheimer} et~al.}{2012}]{oppenheimer12}
{Oppenheimer} B.~D.,  {Dav{\'e}} R.,  {Katz} N.,  {Kollmeier} J.~A.,
  {Weinberg} D.~H.,  2012, \mn@doi [\mnras] {10.1111/j.1365-2966.2011.20096.x},
  \href {https://ui.adsabs.harvard.edu/abs/2012MNRAS.420..829O} {420, 829}

\bibitem[\protect\citeauthoryear{{Parimbelli}, {Branchini}, {Viel},
  {Villaescusa-Navarro}  \& {ZuHone}}{{Parimbelli}
  et~al.}{2022}]{parimbelli_etal22}
{Parimbelli} G.,  {Branchini} E.,  {Viel} M.,  {Villaescusa-Navarro} F.,
  {ZuHone} J.,  2022, arXiv e-prints, \href
  {https://ui.adsabs.harvard.edu/abs/2022arXiv220900657P} {p. arXiv:2209.00657}

\bibitem[\protect\citeauthoryear{{Pillepich} et~al.,}{{Pillepich}
  et~al.}{2018}]{pillepich18}
{Pillepich} A.,  et~al., 2018, \mn@doi [\mnras] {10.1093/mnras/stx2656}, \href
  {https://ui.adsabs.harvard.edu/abs/2018MNRAS.473.4077P} {473, 4077}

\bibitem[\protect\citeauthoryear{{Planck Collaboration} et~al.,}{{Planck
  Collaboration} et~al.}{2016}]{plancketal}
{Planck Collaboration} et~al., 2016, \mn@doi [\aap]
  {10.1051/0004-6361/201525830}, \href
  {https://ui.adsabs.harvard.edu/abs/2016A&A...594A..13P} {594, A13}

\bibitem[\protect\citeauthoryear{{Rahmati}, {Pawlik}, {Rai{\v{c}}evi{\'c}}  \&
  {Schaye}}{{Rahmati} et~al.}{2013}]{rahmati13}
{Rahmati} A.,  {Pawlik} A.~H.,  {Rai{\v{c}}evi{\'c}} M.,   {Schaye} J.,  2013,
  \mn@doi [\mnras] {10.1093/mnras/stt066}, \href
  {https://ui.adsabs.harvard.edu/abs/2013MNRAS.430.2427R} {430, 2427}

\bibitem[\protect\citeauthoryear{{Rahmati}, {Schaye}, {Crain}, {Oppenheimer},
  {Schaller}  \& {Theuns}}{{Rahmati} et~al.}{2016}]{rahmati16}
{Rahmati} A.,  {Schaye} J.,  {Crain} R.~A.,  {Oppenheimer} B.~D.,  {Schaller}
  M.,   {Theuns} T.,  2016, \mn@doi [\mnras] {10.1093/mnras/stw453}, \href
  {http://adsabs.harvard.edu/abs/2016MNRAS.459..310R} {459, 310}

\bibitem[\protect\citeauthoryear{{Schaye} et~al.,}{{Schaye}
  et~al.}{2015}]{schaye15}
{Schaye} J.,  et~al., 2015, \mn@doi [\mnras] {10.1093/mnras/stu2058}, \href
  {https://ui.adsabs.harvard.edu/abs/2015MNRAS.446..521S} {446, 521}

\bibitem[\protect\citeauthoryear{{Shull}, {Smith}  \& {Danforth}}{{Shull}
  et~al.}{2012}]{shull12}
{Shull} J.~M.,  {Smith} B.~D.,   {Danforth} C.~W.,  2012, \mn@doi [\apj]
  {10.1088/0004-637X/759/1/23}, \href
  {https://ui.adsabs.harvard.edu/abs/2012ApJ...759...23S} {759, 23}

\bibitem[\protect\citeauthoryear{{Smith}}{{Smith}}{2020}]{arcus}
{Smith} R.~K.,  2020, in Society of Photo-Optical Instrumentation Engineers
  (SPIE) Conference Series. p. 114442C, \mn@doi{10.1117/12.2576047}

\bibitem[\protect\citeauthoryear{{Smith}, {Hallman}, {Shull}  \&
  {O'Shea}}{{Smith} et~al.}{2011}]{smith11}
{Smith} B.~D.,  {Hallman} E.~J.,  {Shull} J.~M.,   {O'Shea} B.~W.,  2011,
  \mn@doi [\apj] {10.1088/0004-637X/731/1/6}, \href
  {https://ui.adsabs.harvard.edu/abs/2011ApJ...731....6S} {731, 6}

\bibitem[\protect\citeauthoryear{{Sorini}, {Dav{\'e}}, {Cui}  \&
  {Appleby}}{{Sorini} et~al.}{2022}]{sorini21}
{Sorini} D.,  {Dav{\'e}} R.,  {Cui} W.,   {Appleby} S.,  2022, \mn@doi [\mnras]
  {10.1093/mnras/stac2214}, \href
  {https://ui.adsabs.harvard.edu/abs/2022MNRAS.516..883S} {516, 883}

\bibitem[\protect\citeauthoryear{{Springel} et~al.,}{{Springel}
  et~al.}{2005}]{millenium}
{Springel} V.,  et~al., 2005, \mn@doi [\nat] {10.1038/nature03597}, \href
  {https://ui.adsabs.harvard.edu/abs/2005Natur.435..629S} {435, 629}

\bibitem[\protect\citeauthoryear{{Stocke} et~al.,}{{Stocke}
  et~al.}{2014}]{stocke14}
{Stocke} J.~T.,  et~al., 2014, \mn@doi [\apj] {10.1088/0004-637X/791/2/128},
  \href {https://ui.adsabs.harvard.edu/abs/2014ApJ...791..128S} {791, 128}

\bibitem[\protect\citeauthoryear{{Tanimura}, {Aghanim}, {Kolodzig}, {Douspis}
  \& {Malavasi}}{{Tanimura} et~al.}{2020}]{tanimura20}
{Tanimura} H.,  {Aghanim} N.,  {Kolodzig} A.,  {Douspis} M.,   {Malavasi} N.,
  2020, \mn@doi [\aap] {10.1051/0004-6361/202038521}, \href
  {https://ui.adsabs.harvard.edu/abs/2020A&A...643L...2T} {643, L2}

\bibitem[\protect\citeauthoryear{{Theuns}, {Leonard}, {Efstathiou}, {Pearce}
  \& {Thomas}}{{Theuns} et~al.}{1998}]{theuns98}
{Theuns} T.,  {Leonard} A.,  {Efstathiou} G.,  {Pearce} F.~R.,   {Thomas}
  P.~A.,  1998, \mn@doi [\mnras] {10.1046/j.1365-8711.1998.02040.x}, \href
  {https://ui.adsabs.harvard.edu/abs/1998MNRAS.301..478T} {301, 478}

\bibitem[\protect\citeauthoryear{{Tillman}, {Burkhart}, {Tonnesen}, {Bird},
  {Bryan}, {Angl{\'e}s-Alc{\'a}zar}, {Dav{\'e}}  \& {Genel}}{{Tillman}
  et~al.}{2022}]{tillman_etal22}
{Tillman} M.~T.,  {Burkhart} B.,  {Tonnesen} S.,  {Bird} S.,  {Bryan} G.~L.,
  {Angl{\'e}s-Alc{\'a}zar} D.,  {Dav{\'e}} R.,   {Genel} S.,  2022, arXiv
  e-prints, \href {https://ui.adsabs.harvard.edu/abs/2022arXiv221002467T} {p.
  arXiv:2210.02467}

\bibitem[\protect\citeauthoryear{{Tripp}, {Lu}  \& {Savage}}{{Tripp}
  et~al.}{1998}]{tripp98}
{Tripp} T.~M.,  {Lu} L.,   {Savage} B.~D.,  1998, \mn@doi [\apj]
  {10.1086/306397}, \href
  {https://ui.adsabs.harvard.edu/abs/1998ApJ...508..200T} {508, 200}

\bibitem[\protect\citeauthoryear{{Tripp}, {Savage}  \& {Jenkins}}{{Tripp}
  et~al.}{2000}]{tripp00}
{Tripp} T.~M.,  {Savage} B.~D.,   {Jenkins} E.~B.,  2000, \mn@doi [\apjl]
  {10.1086/312644}, \href
  {https://ui.adsabs.harvard.edu/abs/2000ApJ...534L...1T} {534, L1}

\bibitem[\protect\citeauthoryear{{Tripp}, {Sembach}, {Bowen}, {Savage},
  {Jenkins}, {Lehner}  \& {Richter}}{{Tripp} et~al.}{2008}]{tripp08}
{Tripp} T.~M.,  {Sembach} K.~R.,  {Bowen} D.~V.,  {Savage} B.~D.,  {Jenkins}
  E.~B.,  {Lehner} N.,   {Richter} P.,  2008, \mn@doi [\apjs] {10.1086/587486},
  \href {https://ui.adsabs.harvard.edu/abs/2008ApJS..177...39T} {177, 39}

\bibitem[\protect\citeauthoryear{{Verner}, {Verner}  \& {Ferland}}{{Verner}
  et~al.}{1996}]{verner}
{Verner} D.~A.,  {Verner} E.~M.,   {Ferland} G.~J.,  1996, \mn@doi [Atomic Data
  and Nuclear Data Tables] {10.1006/adnd.1996.0018}, \href
  {https://ui.adsabs.harvard.edu/abs/1996ADNDT..64....1V} {64, 1}

\bibitem[\protect\citeauthoryear{{Villaescusa-Navarro}
  et~al.,}{{Villaescusa-Navarro} et~al.}{2022}]{camelsrelease}
{Villaescusa-Navarro} F.,  et~al., 2022, arXiv e-prints, \href
  {https://ui.adsabs.harvard.edu/abs/2022arXiv220101300V} {p. arXiv:2201.01300}

\bibitem[\protect\citeauthoryear{{Werner}, {Finoguenov}, {Kaastra},
  {Simionescu}, {Dietrich}, {Vink}  \& {B{\"o}hringer}}{{Werner}
  et~al.}{2008}]{werner08}
{Werner} N.,  {Finoguenov} A.,  {Kaastra} J.~S.,  {Simionescu} A.,  {Dietrich}
  J.~P.,  {Vink} J.,   {B{\"o}hringer} H.,  2008, \mn@doi [\aap]
  {10.1051/0004-6361:200809599}, \href
  {https://ui.adsabs.harvard.edu/abs/2008A&A...482L..29W} {482, L29}

\bibitem[\protect\citeauthoryear{{Wijers}, {Schaye}, {Oppenheimer}, {Crain}  \&
  {Nicastro}}{{Wijers} et~al.}{2019}]{Wijers}
{Wijers} N.~A.,  {Schaye} J.,  {Oppenheimer} B.~D.,  {Crain} R.~A.,
  {Nicastro} F.,  2019, \mn@doi [\mnras] {10.1093/mnras/stz1762}, \href
  {https://ui.adsabs.harvard.edu/abs/2019MNRAS.488.2947W} {488, 2947}

\makeatother
\end{thebibliography}




\appendix

\section{HI Photoionisation Correction}\label{sec:uv_correction}

The photoionisation corrections are found using the code that generated the Lyman-$\alpha$ spectra released with the CAMELS public data release \citep{camelsrelease}. The spectra were generated using the publicly available code outlined in \citet{Bird:2015} and \citet{Bird:2017}\footnote{\url{https://github.com/sbird/fake_spectra}}. Utilizing addition code from the `fake spectra' package, we can calculate UVB corrections. We recalculate the temperature and electron abundance given a photo correction factor with the hydrogen and helium photoionising and photoheating values and the appropriate recombination values. By solving the ionisation equilibrium equation, we can find the corrected neutral hydrogen fraction and generate new UVB corrected Lyman-$\alpha$ spectra. 

To find the UVB correction factors used in this study, we fit the new corrected Lyman-$\alpha$ spectra to the observed \citet{danforth16} CDDF data. We fit a range of UVB correction factors using a simple $\chi^2$ reduction method. The CDDFs generated from these spectra files are calculated similarly to the method outlined in \citet{Burkhart_2022}. We find these CDDFs by direct integration rather than through Voigt fitting.

We refer the reader to \citet{tillman_etal22} for more details on the comparison of feedback models in cosmological simulations on HI column densities. 

\begin{figure}
\includegraphics[width=0.48\textwidth]{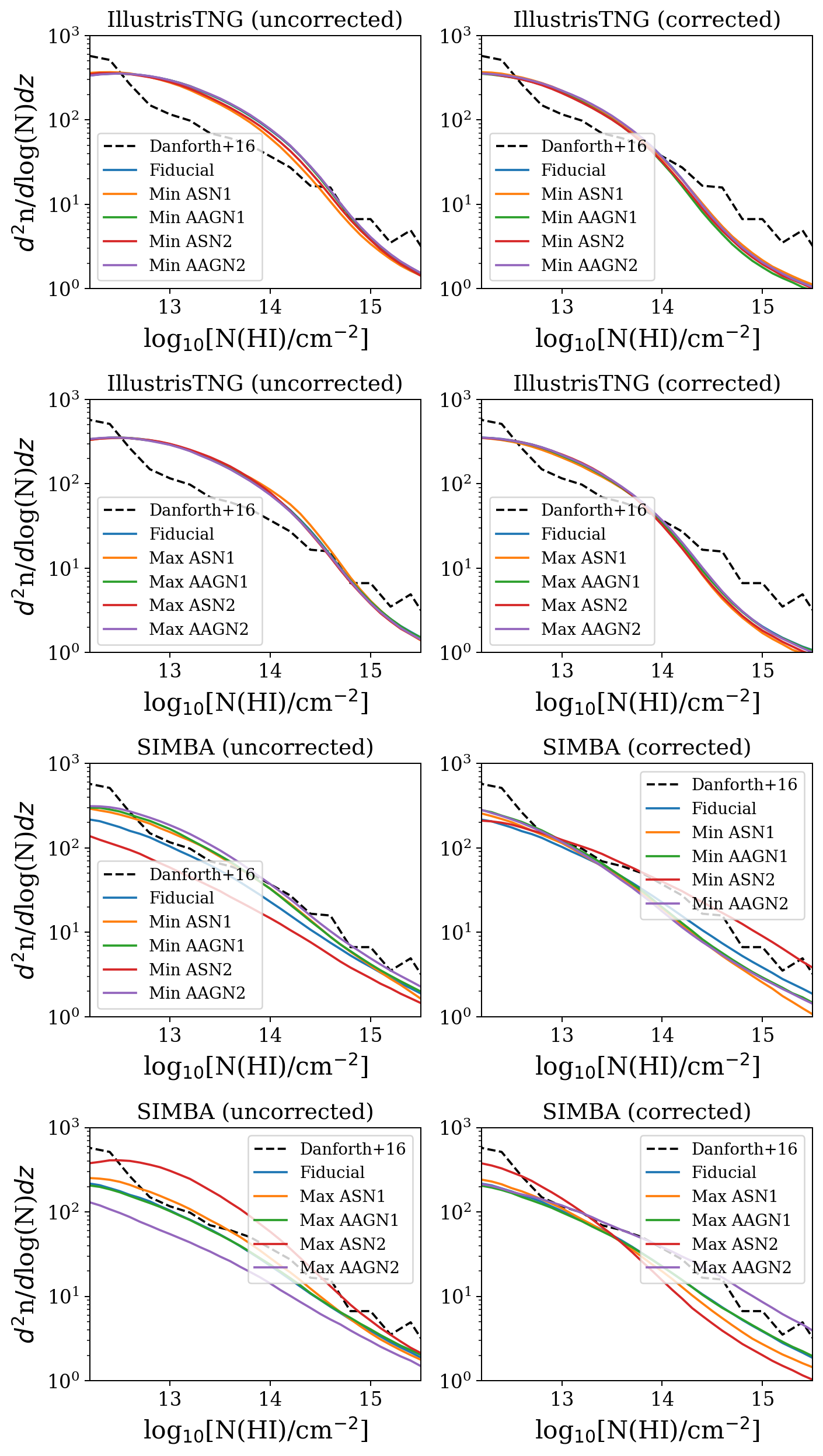}
\caption{CDDFs of HI in the CAMEL simulations. The top four panels show the HI CDDFs of the IllustrisTNG and SIMBA runs with minimum and maximum feedback values for the original simulation outputs. The bottom four show similar plots for the Illustris and SIMBA runs but renormalized using corrections from UV photoionistion, which are detailed in \citet{tillman_etal22}. The dashed lines indicate the observations from \citet{danforth16}.
}
\end{figure}

\section{Box Size and Resolution Convergence}\label{sec:convergence}

Although the CAMEL simulations do not have other box sizes and resolutions, we can use results from previous work to estimate the effects of changing both. For box size, \citet{Wijers} explored EAGLE simulations \citep{schaye15,crain15} with box sizes from 25 to 100 comoving Mpc on a side.  They found deviations of up to 0.2 dexes higher below $N_\mathrm{OVII}=10^{16} {\rm cm}^{-2}$ for a 25~Mpc volume compared to the main 100~Mpc volume, which is the primary EAGLE simulation and contains $64\times$ more volume than the 25~Mpc volume. The CAMELS volumes, at 37.25~Mpc, contain $19\times$ less volume than the primary EAGLE simulation and fall between the 25 and 50 Mpc volumes shown in fig. A3 of \citet{Wijers}. The 50~Mpc EAGLE volume shows much better convergence than the 100~Mpc volume, suggesting that the box size effect is a $\sim 0.1-0.2$ dex change at most.  

As for resolution convergence, \citet{Wijers} explores a factor of $8\times$ higher mass resolution showing better than 0.1 dex convergence below $N_\mathrm{OVII}=10^{15.5} {\rm cm}^{-2}$.  However, the CAMELS gas fluid element resolution is $1.89\times 10^7 {\rm M}_{\odot}$, which is $10\times$ lower than the EAGLE resolution. Therefore, we consider the IllustrisTNG100-2 simulation with a mass resolution of $1.1\times 10^7 {\rm M}_{\odot}$ \citep{nelson19}.  \citet{nelson18} demonstrated this simulation's $N_\mathrm{OVI}$ CDDF is converged to within within 0.1 dex of the $8\times$ better resolution IllustrisTNG100-1 simulation.  While this is a lower oxygen species, we expect the 0.1 dex resolution convergence to be a good indicator for OVII. The CAMELS' smaller box size and lower resolution than other published results from the larger and higher resolution EAGLE and IllustrisTNG main simulation runs suggest that OVII will not change by more than 0.2 dex, which cannot make up for the shortfall compared to the K19 observation.



\bsp	
\label{lastpage}
\end{document}